\documentclass[10pt,a4paper]{article}

\usepackage{times}
\usepackage[centertags]{amsmath}
\usepackage{amssymb,amsfonts,theorem,stmaryrd}
\usepackage{slashbox}
\usepackage{pstricks}

\allowdisplaybreaks

\theorembodyfont{\slshape}
\newtheorem{thm}{Theorem}[section]
\newtheorem{prop}[thm]{Proposition}

\newtheorem{lemma}[thm]{Lemma}
\newtheorem{cor}[thm]{Corollary}
\theorembodyfont{\rmfamily} 
\newtheorem{Def}[thm]{Definition}
\newtheorem{rem}[thm]{Remark}

\newenvironment{pf}{\textit{Proof. }}{\hfill \Box}

\newcommand{\R}{\mathbb{R}}

\newcommand{\Z}{\mathbb{Z}}
\newcommand{\T}{\mathcal{T}}

\newcommand{\HH}{{\mathcal{H}}}
\newcommand{\OO}{\operatorname{O}}
\newcommand{\Riem}{\operatorname{Riem}}
\newcommand{\riem}{\operatorname{riem}}
\newcommand{\W}{\operatorname{weyl}}

\newcommand{\ric}{\operatorname{ric}}

\newcommand{\supp}{\operatorname{supp}}

\newcommand{\id}{\operatorname{id}}
\newcommand{\tr}{\operatorname{tr}}
\newcommand{\Tr}{\operatorname{Tr}}
\newcommand{\rk}{\operatorname{rk}}

\newcommand{\vol}{\operatorname{vol}}

\newcommand{\divergenz}{\operatorname{div}}
\newcommand{\End}{\operatorname{End}}
\newcommand{\im}{\operatorname{im}}

\newcommand{\bb}{\begin{eqnarray}}
\newcommand{\ee}{\end{eqnarray}}
\newcommand{\eee}{\nonumber\end{eqnarray}}

\renewcommand{\Box}{$\boxbox$}

\begin{document}
\title{Chiral Asymmetry and the Spectral Action}
\author{Frank Pf\"aff\-le\footnote{Mathematisches Institut, Georg-August Universit\"at G\"ottingen, Bunsenstra{\ss}e 3-5, G\"ottingen, e-mail: pfaeffle@uni-math.gwdg.de} 
\and Christoph A.~Stephan\footnote{Fachbereich Mathematik, Universit\"at Hamburg, Bundesstra{\ss}e 55, Hamburg,  e-mail: christophstephan@gmx.de}
\footnote{Institut f\"ur Mathematik, Universit\"at Potsdam, Am Neuen Palais 10, Potsdam}   }

\maketitle

\begin{abstract}
\noindent
We consider orthogonal connections with arbitrary torsion on compact Riemannian manifolds.
For the induced Dirac operators, twisted  Dirac operators and Dirac operators of Chamseddine-Connes type we compute the spectral action.
In addition to the Einstein-Hilbert action and the bosonic part of the Standard Model Lagrangian we find the Holst term from Loop Quantum Gravity, a coupling of the Holst term to the scalar curvature and a prediction for the value of the Barbero-Immirzi parameter.
\end{abstract}


\section{Introduction}
In \cite{ConnesChamseddine1} Chamseddine and Connes introduced their spectral action which is motivated by  eigenvalue counting of a Dirac operator.
In the high energy asymptotics the action gives the Einstein-Yang-Mills-Higgs Lagrangian of the Standard Model of particle physics (in its euclidean version) -- given the Dirac operator is suitably chosen.
The geometric setup for this Dirac operator $D_\Phi$ is a $4$-dimensional compact Riemannian spin manifold $M$ without boundary, its spinor bundle $\Sigma M$, a Hermitian vector bundle $\HH$ encoding the particle content and a field $\Phi$ of endomorphisms of $\HH$ containing the Higgs field and the Yukawa matrices.\medskip

\noindent
This operator $D_\Phi$ arises naturally in the framework of noncommutative geometry (see \cite{Connes94} and \cite{ConnesMarcolli}), where spinor fields with coefficients in $\HH$, i.e.~sections of the twisted bundle $\Sigma M\otimes \HH$, find their correspondence in almost-commutative geometries and the endomorphism field $\Phi$ is interpreted as connection of the respective finite space.
\medskip

\noindent
On the tangent bundle and on the spinor bundle we allow more general connections than the Levi-Civita connection, namely we consider orthogonal connections with arbitrary torsion.
By \'E.~Cartan's classification (\cite{Cartan23}, \cite{Cartan24}, \cite{Cartan25}) the torsion then consists of three components: the totally anti-symmetric one, the vectorial one and torsion of Cartan type.
The Cartan type torsion does not contribute to the Dirac operator whereas the other two  components do.
In presence of vectorial torsion the Dirac operator is not symmetric, nevertheless its spectrum is descrete but in general not real-valued.
For the physical consequences of the use of torsion connections in Lorentzian geometry we refer to the classical review \cite{HHKN76} and the more recent overview \cite{S02} and references therein. 
\medskip

\noindent
In this article we consider operators $D_\Phi$ of Chamseddine-Connes type induced by orthogonal connections with torsion, and we permit that the twist bundle $\HH$ may have a chiral asymmetry, i.e.~the number of right-handed particles differs form the number of left-handed ones.
We aim at computing the spectral action for the restriction of $D_\Phi^*D_\Phi$ to sections of the physically relevant subbundle of $\Sigma M\otimes \HH$.
To this end we derive various Lichnerowicz formulas in order to determine the Seeley-deWitt coefficients which appear in the asymptotics of the spectral action.\medskip

\noindent
Our main result is Proposition~\ref{prop_last_Seeley}.
Here the spectral action contains more terms than in the classical situation of \cite{ConnesChamseddine1}.
Some of them vanish if $\HH$ has a chiral symmetry, the other ones are zero for the Levi-Civita connection.
We find the Holst term from Loop Quantum Gravity (\cite{Rovelli},\cite{Thiemann}), a coupling of the Holst term to the scalar curvature and a prediction for the value of the Barbero-Immirzi parameter: 
it turns out that its value depends on the particle content and its absolute value is at least one.
We discuss all these terms in Remarks~\ref{rem_erst}--\ref{rem_letzt}.\medskip 

\noindent
We hope to improve the readability of this text by transfering technicalities such as curvature identies and calculations for characteristic classes into the appendices A--C.
\medskip

\noindent
{\bf Acknoledgement:} The authors appreciate financial support from the SFB 647: {\it Raum-Zeit-Materie}
funded by the Deutsche Forschungsgemeinschaft.
We would like to thank Christian B\"ar, John W.~Barrett and Thomas Sch\"ucker for their support and helpful discussions.

\section{Heat Coefficients for Dirac Operators}
\subsection{Preliminaries and Definitions}
\noindent
Let $M$ be a  manifold of dimension $n\ge 3$, equipped with a Riemannian metric $g$.
Let $\nabla^{g}$ denote the Levi-Civita connection on the tangent bundle $TM$.
We consider an {\it orthogonal} connection $\nabla$ on  $TM$ i.e.\ $\nabla$ is compatible with $g$ in the sense that for all vector fields $X,Y,Z$ one has
$\partial_X \left\langle Y,Z \right\rangle=\left\langle \nabla_XY,Z\right\rangle+\left\langle Y,\nabla_XZ\right\rangle$,
where $\left\langle\cdot,\cdot\right\rangle$ denotes the scalar product given by $g$.
Such connections were first investigated by \'E.~Cartan in \cite{Cartan23,Cartan24,Cartan25}.
In this article we will use the notation from \cite[Sect.~2]{PfaeffleStephan11a}.
It is known (see \cite[Thm.~3.1]{Tricerri}) that orthogonal connections have a specific representation.
To that end one introduces the space of {\it Cartan type torsion tensors} at $p\in M$
\begin{eqnarray*}
\T_3(T_pM)&=& \Big\{  A\in {\bigotimes}^3T^*_pM\;\Big|\; \forall X,Y,Z:\,A(X,Y,Z)+ A(Y,Z,X)+ A(Z,X,Y)=0,\\
& & \qquad\qquad\qquad\qquad\qquad\qquad\qquad\qquad A(X,Y,Z)=-A(X,Z,Y)\;\;
\mbox{ and }\;\; c_{12}(A)(Z)=0\Big\}
\end{eqnarray*}
where $c_{12}$ denotes the trace taken over the first two entries, i.e.\
\[
c_{12}(A)(Z)=\sum_{i=1}^n A(e_i,e_i,Z)
\]
for an orthonormal basis $e_1,\ldots,e_n$ of $T_pM$.

\begin{lemma}[Cor.\ 2.4 in \cite{PfaeffleStephan11a}]\label{Lemma_orthogonal_connections_repr}
For any orthogonal connection $\nabla$  on the tangent bundle of $M$ there exist a unique vector field $V$, a unique $3$-form $T$ and a unique $(3,0)$-tensor field $S$ with $S_p\in \T_3(T_pM)$ for any $p\in M$ such that
for any vector fields $X,Y$ one has
\begin{equation}\label{orthogonal_connection}
\nabla_XY=\nabla^g_XY +\langle X,Y \rangle V- \langle V,Y \rangle X + T(X,Y,\cdot)^\sharp + S(X,Y,\cdot)^\sharp 
\end{equation}
where $T(X,Y,\cdot)^\sharp$ and $S(X,Y,\cdot)^\sharp$ are the unique vectors characterised by $T(X,Y,Z)=\left\langle T(X,Y,\cdot)^\sharp,Z\right\rangle$ and $S(X,Y,Z)=\left\langle S(X,Y,\cdot)^\sharp,Z\right\rangle$ for all $Z$.\hfill \Box
\end{lemma}\medskip

\noindent
For $(k,0)$-tensors $P,Q\in {\bigotimes}^kT^*_pM$ we will use the natural scalar product given by
\begin{equation}\label{tensor_scalar_product}
\langle P,Q \rangle= \sum_{i_1,\ldots,i_k} P(e_{i_1},\ldots,e_{i_k})\cdot Q(e_{i_1},\ldots,e_{i_k})
\end{equation}
where $e_1,\ldots,e_n$ is again an orthonormal basis of $T_pM$.
The induced norm is denoted by $\|P\|$.\medskip

\noindent
The curvature tensors of an orthogonal connection $\nabla$ are defined as
\begin{eqnarray}
\Riem(X,Y)Z&=&\nabla_X\nabla_YZ-\nabla_Y\nabla_XZ -\nabla_{[X,Y]}Z ,\label{def_Riem}\\
\riem(X,Y,Z,W)&=& \langle \Riem(X,Y)Z,W\rangle, \label{def_riem}\\
\ric(X,Y)&=& \sum_{i=1}^n \riem(e_i,X,Y,e_i),\label{def_ric}\\
R &=&\sum_{i=1}^n\ric(e_i,e_i)\label{def_R}.
\end{eqnarray}
For $\nabla=\nabla^g$ we write $\Riem^g$, $\riem^g$, $\ric^g$ and $R^g$, respectively. 
The relevant calculations for norms of these curvature tensors are postponed into Appendix~\ref{appendix_A}.\medskip

\noindent
From now on, let us assume that $M$ carries a spin structure in order to have spinor fields.
In particular, $M$ is then orientable.
Any orthogonal connection $\nabla$ as in \eqref{orthogonal_connection} induces a unique connection on the spinor bundle $\Sigma M$ (see \cite[Chap.~II.4]{Lawson} or \cite[p.\ 17f]{AgricolaSrni}) which is denote by $\nabla$.
Then the Dirac operator associated to $\nabla$ is given by
\begin{equation}\label{def_torsiondirac}
D\psi 
= D^g\psi +\tfrac32 T\cdot \psi -\tfrac{n-1}{2}V\cdot \psi
\end{equation}
where $D^g$ is the Dirac operator associated to the Levi-Civita connection and ``$\cdot$'' is the Clifford multiplication\footnote{For the Clifford relations we use the convention $X\cdot Y+Y\cdot X=-2\,g(X,Y)$ for any tangent vectors $X,Y$, and any $k$-form $\theta^{i_1}\wedge\ldots\wedge\theta^{i_k}$ acts on some spinor $\psi$ by $\theta^{i_1}\wedge\ldots\wedge\theta^{i_k}\cdot\psi = e_{i_1}\cdot\ldots \cdot e_{i_k}\cdot \psi$.} (compare equation (11) in \cite{PfaeffleStephan11b}).
As the Clifford multiplication by the vector field $V$ is skew-adjoint we get that
$D$ is a symmetric operator in the space of  $L^2$-spinor fields if and only if $V\equiv 0$ (see \cite{FriedrichSulanke} and \cite{PfaeffleStephan11a}, and \cite{GS87} for the Lorentzian setting).
Note that the Cartan type torsion $S$ does not contribute to the Dirac operator $D$ (see e.g.~\cite[Lemma~4.7]{PfaeffleStephan11a}).
As $D^*D$ is a generalised Laplacian one has a Lichnerowicz formula.

\begin{thm}[Thm.\ 3.3 in \cite{PfaeffleStephan11b}]
For the Dirac operator $D$ associated to  $\nabla$  we have
\begin{eqnarray}\label{Lichnerowicz}
D^*D\psi &=& \Delta\psi + \tfrac14 R^g\,\psi + \tfrac32 dT\cdot \psi -\tfrac34\|T\|^2\,\psi\nonumber \\
&& \; + \tfrac{n-1}{2}\divergenz^g(V)\,\psi +\left(\tfrac{n-1}{2}\right)^2(2-n)\,|V|^2\,\psi \nonumber \\
&& \; +3(n-1)\Big(T\cdot V\cdot\psi +(V\lrcorner T)\cdot\psi \Big)
\end{eqnarray}
where $\Delta$ is the Laplacian associated to the connection
\[
\widetilde{\nabla}_X\psi = \nabla^g_X\psi +\tfrac32 (X\lrcorner T)\cdot \psi - \tfrac{n-1}{2}V\cdot X\cdot \psi - \tfrac{n-1}{2}\, \langle V, X\rangle\, \psi.
\]
for any spinor field $\psi$.\hfill \Box
\end{thm}

\noindent
We note that the spinor connection $\widetilde{\nabla}$ is induced by the orthogonal connection on the tangent bundle given by
\begin{equation}\label{def_modified_connection}
\widetilde{\nabla}_X Y= \nabla^g_X Y+(n-1)\cdot\left(\langle X,Y \rangle V- \langle V,Y \rangle X\right) + 3\cdot T(X,Y,\cdot)^\sharp.
\end{equation}
It is a modification of $\nabla$ as in \eqref{orthogonal_connection} obtained by replacing $T$ by $3T$, $V$ by $(n-1)V$, and $S$ by zero.\medskip

\noindent
For the case of totally anti-symmetric torsion, i.e.~$V\equiv0$, the above Lichnerowicz formula is due to Agricola and Friedrich (Theorem 6.2 in \cite{AgricolaFriedrich}).

\subsection{Heat Coefficients}
Let $M$ be an compact $n$-dimensional Riemannian manifold without boundary, let $\mathcal{E}\to M$ be Riemannian or Hermitian vector bundle.
Let $H$ be a generalised Laplacian acting on sections in $\mathcal{E}$, i.e.\ the principal symbol of $H$ is given by the Riemannian metric of $M$.
We assume that $H$ is symmetric in the space of $L^2$-sections in $\mathcal{E}$, then the general theory (\cite[Prop.~2.33]{BGV}) states that $H$ is essentially selfadjoint.
Then there exists a unique connection $\nabla^H$ on $\mathcal{E}$ which is compatible with the Riemannian or Hermitian structure, and there is a unique smooth field $E$ of symmetric endomorphisms of $\mathcal{E}$ such that the Bochner formula
\begin{equation}\label{Bochner_formula}
H=\Delta^H- E
\end{equation}
holds, where $\Delta^H$ denotes the Laplacian associated to $\nabla^H$ (see \cite[Prop.\ 2.5]{BGV}).\medskip

\noindent
By elliptic regularity theory $H$ is a smoothing operator, and therefore $\exp(-tH)$ possesses a smooth Schwartz kernel $k_t(x,y)\in \mathcal{E}_x \otimes \mathcal{E}^*_y$ with $t>0$, $x,y\in M$, called the {\it heat kernel} for $H$.
On the diagonal $\{x=y\}\subset M\times M$ the heat kernel has a complete asymptotic expansion, as $t\searrow 0$.

\begin{thm}[Lemma 1.8.2 and Theorem 4.1.6 in \cite{Gilkey95}]\label{Gilkey_asympt}
For any integer $\ell\ge 0$ there is a smooth field $\alpha_{2\ell}(H)$ of endomorphisms of $\mathcal{E}$ such that the short time asymptotics
\[
k_t(x,x)\sim \sum_{\ell\ge 0} t^{\ell-n/2} \alpha_{2\ell}(H)(x)\;\mbox{ as }t\searrow 0
\]
holds uniformly in all $x\in M$.
For $\nabla^H$ and $E$ as in \eqref{Bochner_formula} the first of these endomorphism fields are 
\begin{eqnarray*}
\alpha_0(H)(x) &=& (4\pi)^{-n/2}\;\id_{\mathcal{E}_x},\\
\alpha_2(H)(x) &=& (4\pi)^{-n/2}\;\left( E(x)+\tfrac16\,R^g(x)\,\id_{\mathcal{E}_x}\right),\\
\alpha_4(H)(x) &=& (4\pi)^{-n/2}\; \tfrac{1}{360}\,\Big( 60\,(\Delta^H_\otimes E)(x) + 60\, R^g(x)\,E(x)+180\,E(x)^2+30\,\sum_{i,j=1}^n\Omega_{ij}\Omega_{ij}\\
&&\qquad\qquad\qquad\quad +(12\,(\Delta^g R^g)(x) +5\,(R^g(x))^2-2\,\|\ric^g\|^2+2\,\|\riem^g\|^2)\id_{\mathcal{E}_x} \Big)
\end{eqnarray*}
where $\Omega_{ij}=\nabla^H_{e_i} \nabla^H_{e_j}   -\nabla^H_{e_j}\nabla^H_{e_i}- \nabla^H_{[e_i,e_j]}$ is the 
curvature endomorphism of $\nabla^H$ on the fibre $\mathcal{E}_x$ taken with respect to an orthonormal basis $e_1,\ldots,e_n$ of $T_pM$.
Furthermore $\nabla^H$ induces a connection $\nabla^H_\otimes$ on the endormorphism bundle $\End(\mathcal{E})$, and $\Delta^H_\otimes$ denotes the associated Laplacian.
\hfill\Box
\end{thm}

\noindent
From the asymptotics of the heat kernel on the diagonal one deduces the asymptotics of the heat trace.

\begin{cor}
Let $M$ be a compact Riemannian manifold without boundary, let $\mathcal{E}\to M$ be Riemannian or Hermitian vector bundle.
For a generalised Laplacian $H$ acting on sections in $\mathcal{E}$ we choose $\nabla^H$ and $E$ as in \eqref{Bochner_formula}.
Then one obtains the short time asymptotics of the $L^2$-trace of the heat operator, as $t\searrow 0$,
\begin{equation}\label{eqn_heat_trace_asympt_general}
\Tr_{L^2}\left(e^{-tH} \right)\;\sim\;  \sum_{\ell=0}^\infty t^{\ell-n/2}\, a_{2\ell}(H).
\end{equation}
with $a_{2\ell}(H)=\int_M \tr_{\mathcal{E}_x}\left(\alpha_{2\ell}(H)(x)\right)\,dx $ \hfill \Box
\end{cor}

\noindent
The numbers $a_{2\ell}(H)$ are called the {\it heat coefficients} or the {\it Seeley-deWitt coefficients}.
Theorem~\ref{Gilkey_asympt} gives
\begin{align}
a_0(H)&= (4\pi)^{-n/2}\,\rk(\mathcal{E})\cdot \vol(M),\nonumber\\
a_2(H)&= (4\pi)^{-n/2}\,\int_M \Big(\tr_{\mathcal{E}_x}(E(x))+ \tfrac 16 \,\rk(\mathcal{E})\cdot R^g(x)\Big)\,dx,\label{simple_seeley_dewitt_coefficients} \\
a_4(H)&= (4\pi)^{-n/2} \,\tfrac{1}{360}\,\int_M \Big(60 \,R^g(x)\,\tr_{\mathcal{E}_x}(E(x)) +180 \, \tr_{\mathcal{E}_x}(E(x)^2) +
30 \sum_{i,j=1}^n\tr_{\mathcal{E}_x}\big(\Omega_{ij}\Omega_{ij} \big) \nonumber \\
&\qquad \qquad\qquad\qquad\qquad\qquad\qquad\qquad\quad
+\rk(\mathcal{E})\,\big(5\,(R^g(x))^2- 2\|\ric^g\|^2+ 2\|\riem^g\|^2 \Big)\,dx.\label{seeley_dewitt_coefficients}
\end{align}
For the evaluation of $a_4(H)$ one uses $\tr_{\mathcal{E}_x}((\Delta^H_\otimes E)(x))=\Delta^g( \tr_{\mathcal{E}}E)(x)$ and the fact that $\int_M (\Delta^gf)(x)\,dx=0$ for any smooth function $f$ on $M$. 
\bigskip

\noindent
Now we assume that the bundle $\mathcal{E}$ is $\Z_2$-graded, i.e.\ there is an orthogonal decomposition $\mathcal{E}=\mathcal{E}^+\oplus\mathcal{E}^-$ which is parallel with respect to $\nabla^H_\otimes$. 
This means there exists a parallel field $P$ of endomorphisms of $\mathcal{E}$ such that $P$ is pointwise the orthogonal projection of 
$\mathcal{E}$ onto $\mathcal{E}^+$. 
\medskip

\noindent
In all cases which occur in this article one has $PE = EP$ for the endomorphism $E$ from the Bochner formula \eqref{Bochner_formula}. 
Then $H$ is block diagonal with respect to the decomposition of the space of sections $\Gamma(\mathcal{E})=\Gamma(\mathcal{E}^+)\oplus\Gamma(\mathcal{E}^-)$ and $P$ commutes with $H$, i.e.\ for any section $\Psi\in \Gamma(\mathcal{E})$ we have $PH\Psi=HP\Psi$.
We put $H^+=PH$ and notice that $H^+$ is a generalised Laplacian acting on sections of $\mathcal{E}^+$. Its heat kernel is given by $k^{+}_t(x,y)=P(y)k_t(x,y)$ for $t>0$, $x,y\in M$ and for the corresponding coefficients in the heat kernel asymptotics of Theorem~\ref{Gilkey_asympt} one
has 
\begin{equation}
\label{eqn_chiral_coeff}
\alpha_{2\ell} (H^+) = P \, \alpha_{2\ell} (H)
\end{equation}
pointwise on $M$  and similar formulas for the Seeley-deWitt coefficients in \eqref{eqn_heat_trace_asympt_general}.

\section{Connes' Spectral Action Principle and Chiral Projections}

\noindent
Now we specialise to the case of a 4-dimensional manifold $M$, a Riemannian or Hermitian vector bundle $\mathcal{E} \to M$ and
a generalised Laplacian $H$ acting on sections in $\mathcal{E}$.
Next, we want to consider the Chamseddine-Connes spectral action (see \cite{ConnesChamseddine1}) for the generalised Laplacian $H$.
For $\Lambda >0$ it is defined as
\bb
I_{CC} (H) = {\rm Tr}\, F \left( \tfrac{H}{\Lambda^2} \right) 
\nonumber
\ee
where ${\rm Tr}$ denotes the operator trace over $L^2(\mathcal{E})$ 
and $F:\R^+\to\R^+$ is a smooth cut-off function with support in the interval $[0,+1]$ which is constant near the origin.
Using the heat trace asymptotics one gets an asymptotic expression for $I_{CC}(H)$ as $\Lambda\to\infty$ (see \cite{ConnesChamseddine2} for details):
\begin{equation}
I_{CC}(H) = {\rm Tr} \, F \left( \tfrac{H}{\Lambda^2} \right) =
\Lambda^4 \, F_4 \, a_0 (H) + \Lambda^2 \, F_2 \, a_2(H)
+ \Lambda^0 \, F_0 \, a_4(H) + \OO(\Lambda^{-\infty})
\label{spectral_action}
\end{equation}
with the  first three moments of the cut-off function which are given
by
$F_4 = \int_0^\infty s \cdot F(s) \, ds$, $F_2 = \int_0^\infty F(s) \, ds$ and $F_0 = F(0)$.
These moments are independent of the geometry of the manifold.
\medskip

\noindent
From now on we always assume that $M$ carries a spin structure so that the spinor bundle is defined and so are Dirac operator, twisted or generalised Dirac operators on $M$.
Connes' spectral action principle (\cite{Connes95}, \cite{Connes96}) states that one can extract any action functional of interest in physics from the spectral data of a Dirac operator.\medskip

\noindent
In \cite{ConnesChamseddine1} Chamseddine and Connes have considered an operator $D_\Phi$ which is the sum of a twisted Dirac operator (based on the Levi-Civita connection $\nabla^g$ on $M$) and a zero order term.
Taking $H=(D_\Phi)^2$ in \eqref{spectral_action} they recovered the Einstein-Hilbert action in $a_2(H)$, and in  the remaining terms in \eqref{spectral_action}  they find  the full bosonic Lagrangian of the Standard Model of particle physics.
In particular  $I_{CC}(H)$ produces the correct Higgs potential for electro-weak symmetry breaking.
\medskip

\noindent
In the following we consider various Dirac operators $\mathcal{D}$ induced by orthogonal connections with general torsion as in \eqref{orthogonal_connection}.
We will consider $H=\mathcal{D}^*\mathcal{D}$ (since $\mathcal{D}$ is not selfadjoint in general) and the corresponding Seeley-deWitt coefficients in \eqref{spectral_action}.
If the twist bundle has a chiral asymmetry terms known from Loop Quantum Gravity (\cite{Rovelli}, \cite{Thiemann}) arise.
The situation with purely anti-symmetric torsion has been examined before (in \cite{Torsion}, \cite{IochumLevy}, \cite{PfaeffleStephan11a}), and there also are results on Dirac operators with scalar perturbations (in \cite{Sitarz}).

\subsection{The Bosonic Spectral Action}
First we consider the case where the vector bundle $\mathcal{E}$ is the spinor bundle $\Sigma M$ and the generalised Laplacian  is $H=D^*D$ where $D$ is the Dirac operator as in \eqref{def_torsiondirac}.
The Chamseddine-Connes spectral action of $D^*D$ is determined if one knows the second and the fourth Seeley-deWitt coefficient.
Those can be calculated.
\begin{prop}\label{prop_spec_act_1}
Let $C^g$ denote the Weyl curvature of the Levi-Civita connection (as in Lemma~\ref{Weyl_bleibt_Weyl.}) and let $\chi(M)$ denote the Euler characteristics of $M$, and let $\widetilde{R}$ be the scalar curvature of the modified connection $\widetilde{\nabla}$ (given in \eqref{def_modified_connection}).
Then one has
\begin{align*}
a_2(D^*D)& = -\tfrac{1}{48\,\pi^2}\int_M\widetilde{R}(x)\,dx, \\
a_4(D^*D)& = \tfrac{11}{720}\,\chi(M)-\tfrac{1}{320\,\pi^2}\int_M\|C^g\|^2\,dx
-\tfrac{3}{32\,\pi^2}\int_M\left(\|\delta T\|^2 +\|d(V^\flat)  \|^2\right)\,dx
\end{align*}
\end{prop}
\pf{From \eqref{eq_scalar_curvature} we have 
\[
\widetilde{R}= R^g+18\,\divergenz^g(V)-54\,|V|^2-9\,\|T\|^2.
\]
We obtain $a_2(D^*D)$ inserting \eqref{eq_E_trace} into \eqref{simple_seeley_dewitt_coefficients}.
\medskip

\noindent
We denote the Riemann curvature of $\widetilde{\nabla}$ by $\widetilde{\riem}$
and recall (compare Corollary~4.10 in \cite{PfaeffleStephan11a}) that
\begin{equation}\label{eq_riemriem}
\sum_{i,j}\tr_\Sigma\left(\Omega_{ij}\Omega_{ij} \right)=-\tfrac12\,\left\| \widetilde{\riem}\right\|^2.
\end{equation}
In order to evaluate $a_4(D^*D)$ we insert the trace formulas \eqref{eq_E_trace} and \eqref{eq_E_squared_trace} into 
\eqref{seeley_dewitt_coefficients} and we apply Lemma~\ref{lemma_b1} for $\widetilde{\nabla}$ to obtain
\begin{eqnarray*}
a_4(D^*D) &=& \tfrac{1}{96\,\pi^2} \int_M\left(\tfrac16(R^g)^2-\tfrac{19}{30}\|\ric^g\|^2+\tfrac{2}{15}\|\riem^g\|^2-\tfrac14\|C^g\|^2 \right)\,dx \\
&&\quad +\tfrac{1}{96\,\pi^2} \int_M\left(-9\,\|\delta T\|^2-9\,\|d(V^\flat)\|^2 \right)\,dx.
\end{eqnarray*}
The first integral term on the right hand side equals $\tfrac{11}{720}\,\chi(M)-\tfrac{1}{320\,\pi^2}\int_M\|C^g\|^2\,dx $, see \cite{ConnesChamseddine1} and \cite{ConnesChamseddine2}, and the claim follows.
\hfill \Box}
\medskip

\begin{rem}
In the case of totally anti-symmetric torsion, i.e.~$V\equiv 0$, formulas for these Seeley-deWitt coefficients have been given in \cite{Goldthorpe}, \cite{Obukhov}, \cite{Grensing}, \cite{Torsion}, \cite{IochumLevy}, \cite{PfaeffleStephan11a}.
\end{rem}

\noindent
Next we consider the volume form $\omega^g$ acting on the spinor bundle $\Sigma M$.
Setting $P=P^+=\tfrac12\left(\id_\Sigma+\omega^g \right)$ we have a parallel field of orthogonal projections.
If we now calculate the Seeley-deWitt coefficients of $H^+=P^+D^*D$ we obtain relations to Loop Quantum Gravity.
The {\it Holst term}\footnote{Before \cite{Holst} this density already appeared in \cite{HMS}, a sketch of its history can be found in \cite[Section III.D]{BHN11}} for the modified connection $\widetilde{\nabla}$
is the $4$-form
\begin{equation}
\label{def_Holst_term}
\widetilde{C}_H=18\,\left( dT-\langle T,\ast V^\flat \rangle\omega^g\right),
\end{equation}
see Proposition~2.3 in \cite{PfaeffleStephan11b} and its Erratum.
\begin{prop}\label{prop_spec_act_2}
Let $C^g$ denote the Weyl curvature of the Levi-Civita connection, and let $\widetilde{R}$ be the scalar curvature of the modified connection $\widetilde{\nabla}$.
Denote the Euler characteristics of $M$ by $\chi(M)$ and the first Pontryagin class of $M$ by $p_1(M)$.
Then one has
\begin{align*}
a_2(P^+D^*D)& = -\tfrac{1}{96\,\pi^2}\int_M\left( \widetilde{R}\,\omega^g +\widetilde{C}_H\right) \\
a_4(P^+D^*D)& = \tfrac{11}{1440}\,\chi(M) -\tfrac{1}{96}\,p_1(M)
-\tfrac{1}{640\,\pi^2}\int_M\|C^g\|^2\,dx,\\
&\quad
-\tfrac{3}{64\,\pi^2}\int_M\left(\|\delta T\|^2 +\|d(V^\flat)  \|^2\right)\,dx 
+\tfrac{1}{1152\,\pi^2}\int_M \widetilde{R}\,\widetilde{C}_H.
\end{align*}
\end{prop}

\noindent
\pf{For any $\ell\ge 0$ we have 
\begin{equation*}
a_{2\ell}(P^+D^*D)=\tfrac12 \, a_{2\ell}(D^*D) +
\tfrac12 \int_M \tr_\Sigma\left(\alpha_{2\ell}(D^*D)\,\omega^g\right)\,dx,
\end{equation*}
so we only need to calculate $\int_M \tr_\Sigma\left(\alpha_{2\ell}(D^*D)\,\omega^g\right)\,dx$ for $\ell=1,2$.
\medskip

\noindent
Since $\tr_\Sigma(\omega^g)=0$ we have $(4\pi)^2 \,tr_\Sigma(\alpha_2\,\omega^g)=\tr_\Sigma(E\,\omega^g)$.
From \eqref{eq_E_omega_trace} we get $\tr_\Sigma(E\,\omega^g)\omega^g=-\tfrac13\,\widetilde{C}_H$ and therefore
\[
\int_M \tr_\Sigma\left(\alpha_{2\ell}(D^*D)\,\omega^g\right)\,dx  =-\tfrac{1}{48\,\pi^2} \int_M\widetilde{C}_H.
\]
We use $\tr_\Sigma(\omega^g)=0$ again to obtain
\[
\int_M \tr_\Sigma\left(\alpha_{4}(D^*D)\,\omega^g\right)\,dx
= \tfrac{1}{(4\pi)^2}\,\int_M \Big(\tfrac{1}{6}\,R^g\,\tr_{\Sigma}(E\,\omega^g) +\tfrac12 \, \tr_{\Sigma}(E^2\,\omega^g) +
\tfrac{1}{12} \sum_{i,j=1}^4\tr_{\Sigma}\big(\Omega_{ij}\Omega_{ij} \,\omega^g\big)\Big) \,dx.
\]
The explicit formula for $\Omega_{ij}$ in terms of the Riemann curvature $\widetilde{\riem}$ (Theorem 4.15 in \cite[Chap.~II]{Lawson}) gives
\begin{align*}
\sum_{i,j} \tr_\Sigma\left(\Omega_{ij}\Omega_{ij} \,\omega^g \right)&=
\tfrac{1}{16}\sum_{i,j,k,l,m,n} \widetilde{\riem}_{ijkl}\widetilde{\riem}_{ijmn}
\tr_\Sigma\left(e_ke_le_me_n\,\omega^g \right)
\\
&=\tfrac14\,\langle \widetilde{\riem},\id_{\Lambda^2}\otimes \ast \big(\widetilde{\riem}\big)\rangle
\end{align*}
where we use the notations from Appendix~\ref{appendix_A3}.
By \eqref{eq_pontryagin_class} we get
\begin{equation}\label{eq_omegaomega_pontryagin}
\int_M \sum_{i,j} \tr_\Sigma\left(\Omega_{ij}\Omega_{ij} \,\omega^g \right)\,dx =-4\pi^2\, p_1(M).
\end{equation}
From \eqref{eq_E_squared_omega_trace} we obtain $\tr_\Sigma(E^2\,\omega^g)\omega^g= \tfrac{1}{18}(2R^g+\widetilde{R})\widetilde{C}_H$, 
and with $\tr_\Sigma(E\,\omega^g)\omega^g=-\tfrac13\,\widetilde{C}_H$ we are done.
\hfill \Box}

\noindent
The second Seeley-deWitt coefficient $a_2(P^+D^*D)$ coincides with the Holst action for $\widetilde{\nabla}$ with critical Barbero-Immirzi parameter $-1$, see \cite{PfaeffleStephan11b}.
\medskip

\noindent
It is no surprise that 
one finds the first Pontryagin class and the integral of the derivatives of $T$ and $V$
in the fourth Seeley-deWitt coefficient $a_4(P^+D^*D)$.
The authors did not expect that the integral of the product of the scalar curvature $\widetilde{R}$ and the Holst term $\widetilde{C}_H$ occurs.
The next Lemma shows that $\int_M \widetilde{R}\,\widetilde{C}_H$ is not a topological term.

\begin{lemma}\label{lem_rt}
For any compact oriented $4$-dimensional manifold there exists a Riemannian metric $g$ and a $3$-form $T$ such that $\int_M R^g\,dT\ne 0$.
\end{lemma}
\pf{First we fix some arbitrary (smooth) $3$-form $T$ with support $\supp(T)$ contained in an oriented chart.
For any Riemannian metric $g$ we find a smooth function $f^g$ such that
\[
 dT=f^g\,\sqrt{g}\,dx^1\wedge dx^2\wedge dx^3\wedge dx^4
\]
in this chart.
W.l.o.g.~we assume that $f^g(p)>0$ for one point $p$ in the chart.
Since $\sqrt{g} >0$ we observe that the set $A=\{f^g>0 \}$ does not depend on the choice of $g$.
Next we choose a smooth function $K\in C^\infty(M)$ with the following properties:
on some open set, which is contained in $A$, one has $K>0$, on the set $\supp(T)\setminus A$ one has $K=0$, and there is a point $q\in M\setminus\supp(T)$ with $K(q)<0$.
By a classical result by Kazdan-Warner \cite[Theorem 1.1]{KazdanWarner} one can find a Riemannian metric $g$ on $M$ with $R^g\equiv K$.
For this $g$ one has $\int_M R^g\,dT >0$.
\hfill \Box}
\begin{cor}
Neither $\int_M \widetilde{R}\,\widetilde{C}_H$ nor $-\tfrac{3}{64\,\pi^2}\int_M\left(\|\delta T\|^2 +\|d(V^\flat)  \|^2\right)\,dx 
+\tfrac{1}{1152\,\pi^2}\int_M \widetilde{R}\,\widetilde{C}_H$
(which is the sum of the last three terms in $a_4(P^+D^*D)$)
 are topological invariants. 
In particular, both are non-zero in general.
\end{cor}
\pf{
If one of the above terms were topological its value would be independent of the choice of $T$ and $V$.
Rescaling $T$ and $V$ independently would imply that the term in Lemma~\ref{lem_rt} would always vanish.
\hfill \Box}

\begin{rem}
Using considerations involving the Pontryagin class (see Remark~\ref{rem_B4}) it is possible to express $\int_M\widetilde{R}\,\widetilde{C}_H$ in terms of the Ricci curvature.
\end{rem}

\subsection{Particle Lagrangians}
In the models of particle physics the matter content is encoded in a Hermitian vector bundle $\HH\to M$ equipped with a connection $\nabla^\HH$.
As in the preceeding sections we consider an orthogonal connection $\nabla$ as in \eqref{orthogonal_connection} inducing a connection on the spinor bundle $\Sigma M$, which we denote again by $\nabla$.
The tensor connection on the twisted bundle $\mathcal{E}=\Sigma M\otimes \HH$ induces the Dirac operator $D_\HH$ which is given by
\begin{equation}\label{def_twist_Dirac}
D_\HH(\psi\otimes\chi)=\sum_{i=1}^4\left((e_i\cdot \nabla_{e_i}\psi)\otimes \chi+
(e_i\cdot \psi)\otimes (\nabla^\HH_{e_i}\chi) \right)
\end{equation}
for any positively oriented orthonormal frame $e_1,\ldots, e_4$, any section $\psi$ of $\Sigma M$ and any section $\chi$ of $\HH$.
For further calculations we define the auxiliary orthogonal connection $\nabla^0$ by
\begin{equation}\label{def_nabla_null}
\nabla^0_XY=\nabla^g_XY + T(X,Y,\cdot)^\sharp.
\end{equation}
\begin{prop}
For the twisted Dirac operator $D_\HH$ we get
\begin{equation}\label{eq_twist_Bochner}
D_\HH^*D_\HH (\psi\otimes\chi)= \Delta^{\overline{\nabla}}(\psi\otimes\chi)- (E\psi)\otimes \chi-
\tfrac12\,\sum_{i\ne j}(e_i\cdot e_j\cdot\psi)\otimes\Omega_{ij}^\HH\chi
\end{equation}
where $\Delta^{\overline{\nabla}}$ is the Laplacian associated to the tensor connection
\[
\overline{\nabla}=\widetilde{\nabla}\otimes\id_\HH  + \id_{\Sigma M}\otimes \nabla^\HH
\]
and $\Omega_{ij}^\HH=\nabla^\HH_{e_i}\nabla^\HH_{e_j}-\nabla^\HH_{e_j}\nabla^\HH_{e_i}-\nabla^\HH_{[e_i,e_j]}$ is the curvature of $\nabla^\HH$ and 
$E$ is the potential in the Lichnerowicz formula of the untwisted Dirac operator $D$ as in \eqref{eq_def_potential}.
\end{prop}
\pf{ Given a point $p\in M$, we choose the positively oriented orthonormal frame $e_1,\ldots, e_4$ about $p$  synchronous in $p$ with respect to $\nabla^0$.
The connection $\nabla^0$ has been considered in the Appendix of \cite{Torsion} to which we refer.
One has $\nabla^g_{e_i}e_i=0$ in $p$.
For the twisted Dirac operator we get
\begin{align*}
D_\HH (\psi\otimes \chi)  &= \sum_{i=1}^4 \left((e_i\cdot \nabla^0_{e_i}\psi)\otimes \chi+
(e_i\cdot \psi)\otimes (\nabla^\HH_{e_i}\chi) \right)-\tfrac32 (V\cdot \psi)\otimes\chi,
\\
D_\HH^* (\psi\otimes \chi) &=\sum_{i=1}^4 \left((e_i\cdot \nabla^0_{e_i}\psi)\otimes \chi+
(e_i\cdot \psi)\otimes (\nabla^\HH_{e_i}\chi) \right)+\tfrac32 (V\cdot \psi)\otimes\chi.
\end{align*}
This yields
\begin{align*}
D_\HH^*D_\HH (\psi\otimes\chi)&= (D^*D\psi)\otimes\chi+\sum_{i,j=1}^4(e_j\cdot e_j\cdot\psi)\otimes(\nabla^\HH_{e_j}\nabla^\HH_{e_i}\chi)
-2\sum_{i=1}^4 (\nabla^0_{e_i}\psi)\otimes (\nabla^\HH_{e_i}\chi)\\
&\quad +3\sum_{i=1}^4(V\cdot e_i+\langle V,e_i\rangle)\cdot\psi\otimes \nabla^\HH_{e_i}\chi.
\end{align*}
With \eqref{def_modified_connection} we find
\[
\widetilde{\nabla}_X\psi=\nabla^0_X\psi-(X\lrcorner T)\cdot\psi-\tfrac32 V\cdot X\cdot\psi
-\tfrac32\langle V,X \rangle\psi.
\]
which leads to
\begin{align*}
\Delta^{\overline{\nabla}}(\psi\otimes\chi)&= (\Delta \psi)\otimes \chi -\sum_{i=1}^4 \psi \otimes(\nabla^\HH_{e_i}\nabla^\HH_{e_i}\chi)-2\sum_{i=1}^4 (\nabla^0_{e_i}\psi)\otimes (\nabla^\HH_{e_i}\chi)-2\sum_{i=1}^4 (e_i\lrcorner T)\cdot\psi\otimes \nabla^\HH_{e_i}\chi\\
&\quad
+3\sum_{i=1}^4(V\cdot e_i+\langle V,e_i\rangle)\cdot\psi\otimes \nabla^\HH_{e_i}\chi.
\end{align*}
Taking the difference $D_\HH^*D_\HH-\Delta^{\overline{\nabla}}$ we see that all terms containing $V$ explicitly drop out and we are in a situation as in the Appendix of \cite{Torsion}.
\hfill\Box}
\medskip

\noindent
Here we recall that the curvatures of $\overline{\nabla}$, $\widetilde{\nabla}$ and $\nabla^\HH$ are related as follows (see e.g.~(20) in \cite{Torsion}):
\begin{equation}\label{eq_curvature-addition}
\Omega_{ij}^{\overline{\nabla}}(\psi\otimes\chi)=
(\Omega_{ij}\psi)\otimes\chi +\psi\otimes(\Omega_{ij}^\HH\chi).
\end{equation}
\medskip

\noindent
Let $\Phi$  be a field of selfadjoint endomorphisms of the vector bundle $\HH$.
We call $\Phi$ the {\it Higgs endomorphism}.
Adding a zero order term to the twisted Dirac operator we obtain the {\it Chamseddine-Connes Dirac operator}
\[
D_\Phi(\psi\otimes\chi)=D_\HH(\psi\otimes\chi) +(\omega^g\cdot\psi)\otimes (\Phi\,\chi)
\]
for which one has the following Lichnerowicz formula.

\begin{prop}\label{prop_bochner_chamseddine_connes}
For the  Chamseddine-Connes Dirac operator $D_\Phi$ we get
\begin{equation}\label{eq__chamseddine_connes_Bochner}
D_\Phi^*D_\Phi (\psi\otimes\chi)= \Delta^{\overline{\nabla}}(\psi\otimes\chi)-E_\Phi(\psi\otimes\chi)
\end{equation}
where $\Delta^{\overline{\nabla}}$ is the Laplacian associated to the tensor connection
\[
\overline{\nabla}=\widetilde{\nabla}\otimes\id_\HH  + \id_{\Sigma M}\otimes \nabla^\HH
\]
and the potential $E_\Phi$ is given by
\begin{align}
E_\Phi(\psi\otimes\chi) &=
(E\psi)\otimes \chi+
\tfrac12\,\sum_{i\ne j}(e_i\cdot e_j\cdot\psi)\otimes\Omega_{ij}^\HH\chi
+\sum_i (\omega^g\cdot e_i\cdot\psi)\otimes \big[\nabla^\HH_{e_i},\Phi \big]\chi
\nonumber \\
&\qquad -\psi\otimes(\Phi^2\,\chi) - 3\, (V\cdot \omega^g\cdot\psi)\otimes(\Phi\,\chi).\label{eq__chamseddine_connes_Bochner_potential}
\end{align}
\end{prop}
\pf{ 
The auxiliary connection $\nabla^0$ from \eqref{def_nabla_null} a induces connection on the spinor bundle $\Sigma M$.
By $D^0$ we denote the associated twisted Dirac operator acting on section in $\Sigma M\otimes \HH$.
From \cite{FriedrichSulanke} one can conclude that $D^0$ is self-adjoint.
As in (9) of \cite{Torsion} we have
\begin{equation}\label{dreck1}
\big(D^0\,(\omega^g\otimes\Phi)+(\omega^g\otimes\Phi)D^0\big)(\psi\otimes\chi) = -\sum_i (\omega^g\cdot e_i\cdot\psi)\otimes \big[\nabla^\HH_{e_i},\Phi \big]\chi.
\end{equation}
Furthermore we have the relations 
\begin{equation}\label{dreck2}
D_\HH(\psi\otimes\chi)=D^0(\psi\otimes\chi)-\tfrac32 (V\cdot\psi)\otimes\chi\;\mbox{ and }\;
D_\HH^*(\psi\otimes\chi)=D^0(\psi\otimes\chi)+\tfrac32 (V\cdot\psi)\otimes\chi.
\end{equation}
Combining \eqref{dreck1} and \eqref{dreck2} we get
\begin{equation}\label{dreck3}
\big(D_\HH^*\,(\omega^g\otimes\Phi)+(\omega^g\otimes\Phi)D_\HH\big)(\psi\otimes\chi) = -\sum_i (\omega^g\cdot e_i\cdot\psi)\otimes \big[\nabla^\HH_{e_i},\Phi \big]\chi +3\,(V\cdot\omega^g\cdot\psi)\otimes(\Phi\,\chi).
\end{equation}
From the definition of the Chamseddine-Connes Dirac operator it follows
\begin{equation*}
D_\Phi^*D_\Phi = D_\HH^*  D_\HH+ D_\HH^*\,(\omega^g\otimes\Phi)
+(\omega^g\otimes\Phi)D_\HH + \id_{\Sigma M}\otimes\Phi^2.
\end{equation*}
Into this we insert the Lichnerowicz formula \eqref{eq_twist_Bochner} and equation \eqref{dreck3} and obtain the claim.
\hfill\Box}\medskip

\noindent
In realistic particle models there are fermions of different chirality.
The mathematical description of this is a $\Z_2$-grading of the vector bundle $\HH=\HH^r\oplus\HH^\ell$.
We may have $\dim \HH^r\ne\dim \HH^\ell$, in that case we speak of {\it chiral asymmetry}.
Let $\gamma$ denote the corresponding chirality operator, i.e.~one has $\gamma|_{\HH^r}=\id_{\HH^r}$ and $\gamma|_{\HH^\ell}=-\id_{\HH^\ell}$.
In the following we assume that the endomorphism field $\gamma$ is parallel with respect to $\nabla^\HH$.
\medskip

\noindent
On the other hand the volume form $\omega^g$ induces a $\Z_2$-grading of the spinor bundle $\Sigma M=\Sigma^+ M\oplus \Sigma ^-M$ such that $\omega^g|_{\Sigma^\pm M}=\pm\id_{\Sigma^\pm M}$.
\medskip

\noindent
Motivated by experiment one considers the subbundle $\mathcal{E}^+=(\Sigma^+ M\otimes \HH^r )
\oplus (\Sigma ^-M\otimes \HH^\ell)$ of $\mathcal{E}=\Sigma M\otimes \HH$. 
Then the orthogonal projection field $P:\mathcal{E}\to\mathcal{E}^+$ can be expressed as
\begin{equation}\label{def_phys-projector}
P=\tfrac12 \left(\id_{\Sigma M}\otimes\id_\HH\, +\,\omega^g\otimes\gamma  \right),
\end{equation}
and $P$ is therefore parallel.\medskip

\noindent
Finally we want to determine the asymptotic expression of the Chamseddine-Connes spectral action $I_{CC}(H^+)$ for $H^+=PD_\Phi^*D_\Phi$ as in \eqref{spectral_action}.
To that end we just need to calculate the first three Seeley-deWitt coefficients.
We denote the scalar curvature of the connection $\widetilde{\nabla}$ by $\widetilde{R}$ 
and the Holst term for $\widetilde{\nabla}$ by $\widetilde{C}_H$ as before.
\begin{prop}\label{prop_last_Seeley}
For $H^+=PD_\Phi^*D_\Phi$ one gets
\begin{align*}
a_0(H^+) &= \tfrac{1}{8\,\pi^2} \,\rk (\HH)\,\vol(M) ,\\
a_2(H^+) &= -\tfrac{\rk (\HH)}{96\,\pi^2}\,\int_M\left(\widetilde{R}\omega^g+
\frac{\tr_\HH(\gamma)}{\rk (\HH)} \,\widetilde{C}_H \right) 
-\tfrac{1}{8\,\pi^2}\int_M\tr_\HH(\Phi^2)\,dx, \\
a_4(H^+) &= \tfrac{11\,\rk(\HH)}{1440}\,\chi(M)-\tfrac{\tr_\HH(\gamma)}{96}\,p_1(M)
-\tfrac{\rk(\HH)}{640\,\pi^2}\int_M\|C^g\|^2\,dx
-\tfrac{3\,\rk(\HH)}{64\,\pi^2}\int_M\left( \|\delta T\|^2+\|d(V^\flat)\|^2\right)\,dx\\
&\quad +\tfrac{\tr_\HH(\gamma)}{1152\,\pi^2}\int_M\widetilde{R}\widetilde{C}_H
+\tfrac{1}{16\,\pi^2}\int_M\left( \tr_\HH([\nabla^\HH,\Phi]^2)+\tr_\HH(\Phi^4)
+\tfrac{1}{6}\left( R^g-9\|T\|^2\right)\tr_\HH(\Phi^2)\right)\,dx\\
&\quad +\tfrac{1}{96\,\pi^2}\int_M \tr_\HH(\Phi^2\gamma)\,\widetilde{C}_H
+\tfrac{5}{192\,\pi^2}\int_M\tr_\HH(\Omega^\HH\Omega^\HH)\,dx
+\tfrac{1}{64\,\pi^2}\int_M\tr_\HH(\ast\Omega^\HH\Omega^\HH\gamma)\,dx
\end{align*}
with the abbreviations $[\nabla^\HH,\Phi]^2=\sum_i [\nabla^\HH_{e_i},\Phi][\nabla^\HH_{e_i},\Phi] $, $\Omega^\HH\Omega^\HH=\sum_{i,j}\Omega^\HH_{ij}\Omega^\HH_{ij}$ and $\ast\Omega^\HH\Omega^\HH=\sum_{i,j}(\ast\Omega^\HH_{ij})\Omega^\HH_{ij}$.
\end{prop}
\pf{
For all following calculations we use \eqref{eqn_chiral_coeff}.
We note that $\tr_{\mathcal{E}}(\omega^g\otimes\gamma)=\tr_\Sigma(\omega^g)\tr_\HH(\gamma)=0$ and get
\[
 a_0(H^+)=\tfrac{1}{(4\,\pi)^2}\int_M \tr_{\mathcal{E}}(P)\, dx
=\tfrac{1}{8\,\pi^2} \,\rk (\HH)\,\vol(M).
\]
Now we make the standard observation that for $1\le k\le 3$ and $i_1,\ldots,i_k$ pairwise distinct one has
that $\tr_\Sigma(e_{i_1}\cdots e_{i_k})=\tr_\Sigma(\omega^g e_{i_1}\cdots e_{i_k})=0$.
For the traces of the potential $E_\Phi$ from \eqref{eq__chamseddine_connes_Bochner_potential} we then get
\begin{align}
\tr_{\mathcal{E}} (E_\Phi)&= \tr_\Sigma(E)\,\rk(\HH)-4\,\tr_\HH(\Phi^2),\label{eq_trace_EPhi}\\
\tr_{\mathcal{E}} (E_\Phi(\omega^g\otimes\gamma)&=\tr_\Sigma(E\,\omega^g)\tr_\HH(\gamma).
\label{eq_trace_EPhi_omega}
\end{align}
Hence the second Seeley-deWitt coefficient is given by
\begin{align*}
a_2(H^+) &=
\tfrac{1}{32\,\pi^2}\int_M\left(\tr_{\mathcal{E}} (E_\Phi)+\tfrac16 \, R^g\,\rk(\mathcal{E}) \right)dx +
\tfrac{1}{32\,\pi^2}\int_M\left(\tr_{\mathcal{E}} (E_\Phi(\omega^g\otimes\gamma)
 +\tfrac16 \, R^g \tr_{\mathcal{E}}(\omega^g\otimes\gamma)\right)dx \\
&= 
\tfrac{1}{32\,\pi^2}
\int_M\left(\rk(\HH)\,\tr_{\Sigma} (E)+\tfrac23 \,\rk(\HH)\, R^g -4\tr_\HH(\Phi^2)\right)dx +
\tfrac{\tr_\HH(\gamma)}{32\,\pi^2}\int_M\tr_\Sigma(E\,\omega^g) dx.
\end{align*}
Now we insert \eqref{eq_E_trace} and \eqref{eq_E_omega_trace} and obtain the formula for $a_2(H^+)$ in the Proposition.\medskip

\noindent
To evaluate $a_4(H^+)$ we use Theorem~\ref{Gilkey_asympt} and \eqref{eqn_chiral_coeff} and obtain
\begin{align}
a_4(H^+)&= \tfrac{1}{(4\pi)^2}\tfrac{1}{360}\int_M
\Big( 30\,R^g
\big(\tr_{\mathcal{E}}(E_\Phi)+\tr_{\mathcal{E}}(E_\Phi(\omega^g\otimes\gamma) )\big)
+90\,
\big(\tr_{\mathcal{E}}(E_\Phi^2)+\tr_{\mathcal{E}}(E_\Phi^2(\omega^g\otimes\gamma) )\big)
\nonumber \\
&\qquad\qquad\qquad \quad \;\;
+15\,\sum_{i,j}\tr_{\mathcal{E}}\big(\Omega_{ij}^{\overline{\nabla}} \Omega_{ij}^{\overline{\nabla}} \big)
+15\,\sum_{i,j}\tr_{\mathcal{E}}\big(\Omega_{ij}^{\overline{\nabla}} \Omega_{ij}^{\overline{\nabla}}(\omega^g\otimes\gamma) \big)\nonumber \\
&\qquad\qquad\qquad \quad \;\;
+ \,\tr_{\mathcal{E}}(P)\;\big(5(R^g)^2-2\|\ric^g \|^2 +2\|\riem^g\|^2 \big)\,\Big)\,dx\label{superequation}
\end{align}
By means of \eqref{eq_curvature-addition} we identify the terms containing $\Omega_{ij}^{\overline{\nabla}}$
as
\begin{align}
\tr_{\mathcal{E}}\big(\sum_{i,j}\Omega_{ij}^{\overline{\nabla}} \Omega_{ij}^{\overline{\nabla}} \big)
&=\rk(\HH)\,\tr_\Sigma\big(\sum_{i,j}\Omega_{ij}\Omega_{ij}\big)
+4\,\tr_\HH\big(\sum_{i,j}\Omega_{ij}^\HH\Omega_{ij}^\HH\big)\label{eq_omega_nabla_bar}\\
\tr_{\mathcal{E}}\big(\sum_{i,j}\Omega_{ij}^{\overline{\nabla}} \Omega_{ij}^{\overline{\nabla}}(\omega^g\otimes\gamma) \big)
&= \tr_\HH(\gamma)\tr_\Sigma\big(\sum_{i,j}\Omega_{ij}\Omega_{ij}\omega^g\big)\label{eq_ast_omega_nabla_bar}
\end{align}
where we have used  \eqref{eq_curvature-addition} and $\tr_\Sigma(\omega^g)=0$.
In order to evaluate the traces that involve $E_\Phi^2$ one has to write down more than two dozen of terms.
Then one has to check that in the trace most of these terms vanish due to the Clifford relations and the cyclicity of the trace, similarly as in \cite[equations (11) and (12)]{Torsion}.
The traces are
\begin{align}
\tr_{\mathcal{E}}(E_\Phi^2) &= 
\rk(\HH)\,\tr_\Sigma(E^2)+\tr_\HH\big( \sum_{i,j}\Omega_{ij}^\HH\Omega_{ij}^\HH\big)
+\tr_\HH\big(\sum_i[\nabla^\HH_{e_i},\Phi ]^2 \big)
+4\,\tr_\HH(\Phi^4) \nonumber\\
&\quad
-2\tr_\Sigma(E)\,\tr_\HH(\Phi^2)
+36\,|V|^2\,\tr_\HH(\Phi^2)+12\,\partial_V\tr_\HH(\Phi^2),
\label{eq_trace_EPhi_squared}
\\
\tr_{\mathcal{E}}(E_\Phi^2(\omega^g\otimes\gamma) )&=
\tr_\HH(\gamma)\,\tr_\Sigma(E^2\,\omega^g)
+\tr_\HH\big( \sum_{i,j}(\ast\Omega_{ij}^\HH)\Omega_{ij}^\HH\,\gamma\big)
-2\tr_\Sigma(E\,\omega^g)\,\tr_\HH(\Phi^2\,\gamma).\label{eq_trace_EPhi_gamma_squared}
\end{align}
Integration by parts gives $\int_M\partial_V\tr_\HH(\Phi^2)\,dx=-\int_M\divergenz^g(V)\,\tr_\HH(\Phi^2)\,dx$.
We integrate the last three summands in \eqref{eq_trace_EPhi_squared} and find with \eqref{eq_E_trace} that
\begin{equation}\label{mirakuloeseKanzellation}
\int_M\Big( -2\tr_\Sigma(E)+36|V|^2-12\divergenz^g(V)\Big)\,\tr_\HH\big(\Phi^2\big)\,dx= 2\int_M\big(R^g-3\|T\|^2 \big)\,\tr_\HH\big(\Phi^2\big)\, dx.
\end{equation}
To conclude we insert \eqref{eq_trace_EPhi}, \eqref{eq_trace_EPhi_omega} and equations \eqref{eq_omega_nabla_bar}--\eqref{mirakuloeseKanzellation} into \eqref{superequation}
and observe that the occuring terms can be arranged in following way:
\begin{align*}
a_4(H^+)&= \tfrac{\rk(\HH)}{2}\int_M \tr_\Sigma\big(\alpha_4(D^*D) \big)\,dx
+\tfrac{\tr_\HH(\gamma)}{2}\int_M \tr_\Sigma\big(\alpha_4(D^*D) \,\omega^g\big)\,dx\\
&\quad +\tfrac{1}{16\,\pi^2}\int_M\left( \tr_\HH([\nabla^\HH,\Phi]^2)+\tr_\HH(\Phi^4)
+\tfrac{1}{6}\left( R^g-9\|T\|^2\right)\tr_\HH(\Phi^2)-\tfrac12
\,\tr_\Sigma(E\,\omega^g)\,\tr_\HH(\Phi^2\,\gamma)\right)\,dx\\
&\quad
+\tfrac{5}{192\,\pi^2}\int_M\tr_\HH(\Omega^\HH\Omega^\HH)\,dx
+\tfrac{1}{64\,\pi^2}\int_M\tr_\HH(\ast\Omega^\HH\Omega^\HH\gamma)\,dx.
\end{align*}
The terms $\int_M \tr_\Sigma\big(\alpha_4(D^*D) \big)\,dx$ and $\int_M \tr_\Sigma\big(\alpha_4(D^*D) \,\omega^g\big)\,dx$ have already been computed in Propositions~\ref{prop_spec_act_1} and \ref{prop_spec_act_2}.
To simplify the term $\tr_\Sigma(E\,\omega^g)$ we take equation \eqref{eq_E_omega_trace} into account.
\hfill\Box}
\medskip

\noindent
In the following remarks we want to discuss the terms in $a_2(H^+)$ and $a_4(H^+)$ which seem interesting to us.

\begin{rem}\label{rem_erst}
a) In $a_4(H^+)$ one recovers the Yang-Mills functional $\int_M\tr_\HH(\Omega^\HH\Omega^\HH)\,dx$. 
For the Standard Model of particle physics the details have already been given by Chamseddine and Connes in \cite{ConnesChamseddine1}, see also \cite{IochumKastler}.\\
b) If $\HH$ is an associated bundle the term $\int_M\tr_\HH(\ast\Omega^\HH\Omega^\HH\gamma)\,dx$ is topolgical and can be expressed by  Chern classes (compare (4.66) in \cite{Lizzi}).
Details for the Standard Model can be found in \cite{ConnesChamseddine2}, equation (7.5) and the discussion thereafter.
\end{rem}

\begin{rem}
a) The roles of $V$ and $T$ seem dual to each other in the dynamical term $\int_M\left( \|\delta T\|^2+\|d(V^\flat)\|^2\right)\,dx$.
But by the term $\int_M\left( R^g-9\|T\|^2\right)\tr_\HH(\Phi^2)\,dx$
only the anti-symmetric torsion component $T$ couples to the Higgs endomorphism.\\
b) If the relation $\gamma\Phi=-\Phi\gamma$ holds, which is the case in models based on the axioms of noncommutative geometry (\cite{Connes94}, \cite{ConnesChamseddine1}, \cite{ConnesChamseddine2})
 the trace $\tr_\HH(\Phi^2\gamma)$ is zero and
the term $\int_M \tr_\HH(\Phi^2\gamma)\,\widetilde{C}_H$ vanishes.
Otherwise the vectorial torsion component $V$ couples to $\Phi$ (via the Holst term $\widetilde{C}_H$).
\end{rem}

\begin{rem}\label{rem_letzt}
Up to a multiplicative constant the term
$\int_M\big(\widetilde{R}\omega^g+\tfrac{\tr_\HH(\gamma)}{\rk (\HH)} \,\widetilde{C}_H \big)$
coincides with the Holst action for the modified connection $\widetilde{\nabla}$ with Barbero-Immirzi parameter $-\tfrac{\rk (\HH)}{\tr_\HH(\gamma)}$ (whose modulus is larger or equal to 1).
A different derivation for the numerical value of the Barbero-Immirzi parameter in a similar situation can be found in \cite{Broda}.
\end{rem}

\appendix

\section{Norms and Decomposions of Curvature Tensors}\label{appendix_A}
\subsection{The Kulkarni-Nomizu Product}\label{appendix_A_1}
In this section let $U$ denote an $n$-dimensional real vector space, equipped with a scalar product $g$.
\begin{Def}\label{def_Kulkarni_Nomizu}
For two $(2,0)$-tensors $h$ and $k$ on $U$ the {\it Kulkarni-Nomizu product} is the $(4,0)$-tensor $h\owedge k$ given by
\[
(h\owedge k)(X,Y,Z,W)=h(X,Z)k(Y,W)+h(Y,W)k(X,Z)-h(X,W)k(Y,Z)-h(Y,Z)k(X,W)
\]
for any $X,Y,Z,W\in U$.
\end{Def}
We would like to point out that we do not require $h$ and $k$ to be symmetric.
This deviates from the usual definition, see e.g.\ \cite[Def.\ 1.110]{Besse}.
In any case we have $h\owedge k = k\owedge h$.\medskip

\noindent
Following the usual conventions $S^2U^*$ denotes the space of symmetric $(2,0)$-tensors on $U$, and $\Lambda^2 U^*$ is the space of anti-symmetric $(2,0)$-tensors.\medskip

\noindent
We recall that for any $(2,0)$-tensor $k$ on $U$ the {\it trace} (or {\it $g$-trace}) is given by
\[
\tr_g(k)=\sum_{i=1}^n k(E_i,E_i)
\]
for any basis $E_1,\ldots,E_n$ of $U$ which is orthonormal w.r.t.\ $g$.
For a $(4,0)$-tensor $Q$ the {\it Ricci contraction} $c(Q)$ is the $(2,0)$-tensor 
\[
c(Q)(X,Y)=\sum_{i=1}^n Q(E_i,X,Y,E_i)
\]
for any $g$-orthonormal basis $E_1,\ldots,E_n$ of $U$ and any $X,Y\in U$.\medskip

\noindent
Elementary calculations show:
\begin{lemma}\label{lemma_kulkarni_trace_and_norm}
Let $k$ and $h$ be a $(2,0)$-tensors on $U$. 
Then one has
\begin{enumerate}
\item[a)] The Ricci contraction of the Kulkarni-Nomizu product is
$c(k\owedge g)=(2-n)\, k-\tr_g(k)\,g$.
\item[b)] $\langle k\owedge g, h\owedge g \rangle = 4(n-2)\, \langle k,h \rangle + 4\,\tr_g(k)\cdot\tr_g(h)$, where $\langle,\rangle$ denotes the natural scalar product for $(k,0)$-tensors as in \eqref{tensor_scalar_product}.
In particular one obtains for the norms $\|k\owedge g \|^2= 4(n-2)\|k\|^2+ 4(\tr_g(k))^2$ and $\|g\owedge g \|^2= 8n(n-1)$.
\end{enumerate}
\end{lemma}
\noindent
The {\it Bianchi map} $b:\bigotimes^4 U^*\to\bigotimes^4 U^*$ is an endomorphism of the space of $(4,0)$-tensors, mapping $Q\mapsto b(Q)$ with
\[
b(Q)(X,Y,Z,W)= \tfrac13\big( Q(X,Y,Z,W)+ Q(Y,Z,X,W)+Q(Z,X,Y,W)\big)\mbox{ for any }X,Y,Z,W\in U.
\]
It is known (\cite[1.107]{Besse}) that $b$ is an idempotent, and one easily checks that $b$ is symmetric w.r.t.\ $\langle,\rangle$.
Therefore $b$ is an orthogonal projection.\medskip

\noindent
Let $S^2(\Lambda^2 U^*)$ denote the space of $(4,0)$-tensors on $U$ that are anti-symmetric in the first and the last pair of their entries and with the symmetry
$Q(X,Y,Z,W)=Q(Z,W,X,Y)$ for any $X,Y,Z,W\in U$.\medskip

\noindent
Note that $b$ leaves $S^2(\Lambda^2 U^*)$ invariant, thus $S^2(\Lambda^2 U^*)$ orthogonally decomposes into the kernel and the image of $b$.
One calls
\begin{equation}\label{def_algebraic_curvature_tensor}
\mathcal{R}(U^*)= S^2(\Lambda^2 U^*) \cap \ker (b)
\end{equation}
the space of {\it algebraic curvature tensors}, compare \cite[1.108]{Besse},
furthermore one has $S^2(\Lambda^2 U^*) \cap \im (b)=\Lambda^4 U^* $.
\begin{rem}\label{rem_automatisch_curvature}
In general for an orthogonal connections the associated curvature tensor $\riem$ is not an algebraic curvature tensor, it is not even contained in 
$S^2(\Lambda^2 U^*)$. 
\end{rem}
\begin{rem}
For $k,h\in S^2U^*$ we have $k\owedge h\in \mathcal{R}(U^*)$.
\end{rem}\medskip

\noindent
According to the conventions of \cite{Besse} one defines $S^2_0U^*=\left\{h\in S^2U^*\, \big|\,\tr_g(h)=0 \right\}$.\medskip

\noindent
The following classical result is called the Ricci decomposition of an algebraic curvature tensor, see (1.116) in \cite{Besse}.
For $Q\in\mathcal{R}(U)$ the {\it algebraic scalar curvature} is $s(Q)=\tr_g\big(c(Q) \big)\in\R$ and the {\it algebraic tracefree Ricci curvature} is
$h(Q)=c(Q)-\tfrac1n s(Q)\,g\in S^2_0 U^*$.
The {\it algebraic Weyl tensor} $\W(Q)$ is defined as
\[
\W(Q)= Q + \tfrac{1}{n-2}\, h(Q)\owedge g +\tfrac{1}{2n(n-1)} \,s(Q)\cdot g\owedge g.
\]
Then the three summands in the decomposition
\begin{equation}\label{eq_Ricci_decomposition}
Q = -\tfrac{1}{2n(n-1)} \,s(Q)\cdot g\owedge g \, - \, \tfrac{1}{n-2}\, h(Q)\owedge g \, +\,\W(Q)
\end{equation}
are orthogonal with respect to the natural scalar product for $(4,0)$-tensors given as in \eqref{tensor_scalar_product}.

\begin{rem}\label{rem_weyl_orthogonal}
An important property of the algebraic Weyl tensor $\W(Q)$ is that it is tracefree in any pair of entries.
In particular, for any $(2,0)$-tensor $k$ it follows that the $(4,0)$-tensors $\W(Q)$ and $k\owedge g$ are perpendicular.
\end{rem}

\begin{rem}
The algebraic Weyl tensor vanishes, $\W(Q)=0$, if $n\le 3$, and the algebraic tracefree Ricci curvature $h(Q)$ is zero if $n\le 2$.
\end{rem}

\noindent
Finally, let $\Lambda^2(\Lambda^2 U^*)$ denote the space of $(4,0)$-tensors on $U$ that are anti-symmetric in the first and the last pair of their entries and with the symmetry $Q(X,Y,Z,W)=-Q(Z,W,X,Y)$ for any $X,Y,Z,W\in U$.\medskip

\noindent
For any $k\in \Lambda^2U^*$ we note that $k\owedge g\in \Lambda^2(\Lambda^2 U^*)$.
If we take $k=c(Q)$ for $Q\in\Lambda^2(\Lambda^2 U^*)$ we can give an orthogonal decomposition similar to \eqref{eq_Ricci_decomposition}:
\begin{lemma}
For any $Q\in\Lambda^2(\Lambda^2 U^*)$ the two summands in the decomposition
\begin{equation}\label{eq_anti_Ricci_decomposition}
Q =  - \, \tfrac{1}{n-2}\, c(Q)\owedge g \, +\,\big(Q+\tfrac{1}{n-2}\, c(Q)\owedge g \big)
\end{equation}
are orthogonal with respect to the natural scalar product for $(4,0)$-tensors.
\end{lemma}
\pf{ For $Q,P \in\Lambda^2(\Lambda^2 U^*)$  one verifies $c\big(c(Q)\owedge g \big) =(2-n) \,c(Q)$ and $\langle P,c(Q)\owedge g\rangle =\langle c(P)\owedge g,Q\rangle$.
Hence the map $\pi: \Lambda^2(\Lambda^2 U^*)\to \Lambda^2(\Lambda^2 U^*), Q\mapsto \frac{1}{2-n}c(Q)\owedge g$ is a symmetric idempotent and therefore an orthogonal projection. 
Then \eqref{eq_anti_Ricci_decomposition} is just the orthogonal decomposition into image and kernel of $\pi$.
\hfill \Box}\medskip

\noindent
We summerise our consideration so far and a decomposition of $\bigotimes^2(\Lambda^2U^*)=S^2(\Lambda^2U^*)\oplus \Lambda^2(\Lambda^2U^*)$.
We introduce the following vector spaces of tensors:
\begin{align*}
\mathcal{S}(U^*) &= \R g\owedge g ,\\
\mathcal{H}^S(U^*)&= S^2_0(U^*)\owedge g ,\\
\mathcal{H}^A(U^*)&= \Lambda^2(U^*)\owedge g ,\\
\mathcal{W}^S(U^*)&= \ker(c\big|_{\mathcal{R}(U^*)}),\\
\mathcal{W}^A(U^*)&= \ker(c\big|_{\Lambda^2(\Lambda^2U^*)}),\\
\Lambda^4(U^*)&= b\big(S^2(\Lambda^2U^*)\big).
\end{align*}
\begin{prop}\label{prop_riem_orth_decomp}
The decomposition
$
\bigotimes^2(\Lambda^2U^*)= \mathcal{S}(U^*)\oplus \mathcal{H}^S(U^*) \oplus \mathcal{W}^S(U^*)\oplus \Lambda^4(U^*) \oplus \mathcal{H}^A(U^*)\oplus \mathcal{W}^A(U^*)
$
is orthogonal with respect to the natural scalar product for $(4,0)$-tensors given as in \eqref{tensor_scalar_product}.
\end{prop}
\pf{This is a direct consequence of the above considerations.
\hfill\Box}

\begin{rem}\label{rem_tensor_dimensions_if_dim_4}
If $U$ is $4$-dimensional we have
$\dim \mathcal{S}(U^*)= \dim \Lambda^4(U^*)=1$, $\dim \mathcal{H}^S(U^*)=\dim \mathcal{W}^A(U^*)= 9$, $\dim \mathcal{W}^S(U^*)=10$ and $\dim \mathcal{H}^A(U^*)=6$.
\end{rem}\medskip

\subsection{The Hodge $\ast$-Operator}\label{appendix_A3}
Now let $U$ denote a $4$-dimensional real vector space, equipped with a scalar product $g$ and an orientation.
We denote the induced volume form by $\omega^g$.
\medskip

\noindent
Throughout this section we fix an oriented orthonormal basis $E_1,\ldots,E_4$ of $U$ and for any $(4,0)$-tensor $Q$ and any $(2,0)$-tensor $h$ we denote
\[
Q_{ijk\ell}=Q(E_i,E_j,E_k,E_\ell) \mbox{ and } h_{ij}=h(E_i,E_j).
\]
Furthermore $\omega^g(E_i,E_j,E_k,E_\ell)=\epsilon_{ijk\ell}$ and $g(E_i,E_j)=\delta_{ij}$.
\medskip

\noindent
The Hodge $\ast$-operator is an endomorphism of $\Lambda^2U^*$ with $\ast\ast=\id_{\Lambda^2}$.
It naturally induces an endomorphism
$\id_{\Lambda^2}\otimes \ast: \bigotimes^2(\Lambda^2U^*)\to  \bigotimes^2(\Lambda^2U^*)$.

\begin{lemma}
For any $P,Q\in \bigotimes^2(\Lambda^2U^*)$ we have $\langle P ,\id_{\Lambda^2}\otimes \ast(Q)\rangle =\langle\id_{\Lambda^2}\otimes \ast(P) ,Q\rangle$.
In other words $\id_{\Lambda^2}\otimes \ast$ is a selfadjoint endomorphism of $\bigotimes^2(\Lambda^2U^*)\subset \bigotimes^4U^*$.
\end{lemma}
\pf{ Note that $\langle P ,\id_{\Lambda^2}\otimes \ast(Q)\rangle = \sum P_{ijk\ell}Q_{ijab}\epsilon_{abk\ell}= \langle\id_{\Lambda^2}\otimes \ast(P) ,Q\rangle$.
\hfill \Box}

\begin{cor}
$\id_{\Lambda^2}\otimes \ast$ is an isometry of $\bigotimes^2(\Lambda^2U^*)$
as $(\id_{\Lambda^2}\otimes \ast)^2=\id_{\Lambda^2}\otimes\id_{\Lambda^2}$.
\end{cor}

\begin{lemma}\label{lemma_A11}
Let $h$ and $k$ be a $(2,0)$-tensors on $U$. Then we have:
\begin{enumerate}
\item[a)] $\id_{\Lambda^2}\otimes \ast(g\owedge g)=4\,\omega^g$.
\item[b)] If either of $h$ and $k$ is symmetric, then $\langle h\owedge g ,\id_{\Lambda^2}\otimes \ast(k\owedge g )\rangle =0$.
\item[c)] If $h$ and $k$ are both anti-symmetric, then $\langle h\owedge g ,\id_{\Lambda^2}\otimes \ast(k\owedge g )\rangle = 8\langle h,\ast k \rangle$.
\end{enumerate}
\end{lemma}
\pf{
For a) we just note that $\id_{\Lambda^2}\otimes \ast(g\owedge g)_{ij\ell m}=2\sum(\delta_{ia}\delta_{jb}-\delta_{ja}\delta_{ib})\epsilon_{ab\ell m}= 4\,\epsilon_{ij\ell m}$.
For arbitrary $(2,0)$-tensors $h$ and $k$ we calculate
\[
\langle h\owedge g ,\id_{\Lambda^2}\otimes \ast(k\owedge g )\rangle = 8 \sum_{i,j,\ell, m} h_{ij}k_{\ell m}\epsilon_{ij\ell m},
\]
from which b) and c) follow.
\hfill \Box}\medskip

\noindent
In the following proposition we study to what extent the map $\id_{\Lambda^2}\otimes \ast$  respects the decomposition from Proposition~\ref{prop_riem_orth_decomp}.
The relations which we prove have been found before by Hehl et al.\ in \cite[Formula (B.4.35)]{HMMN95}.
\begin{prop}\label{prop_decompo_Hodge}
If we restrict the isometry $\id_{\Lambda^2}\otimes \ast: \bigotimes^2(\Lambda^2U^*)\to  \bigotimes^2(\Lambda^2U^*)$ to the components of the decomposition of $\bigotimes^2(\Lambda^2U^*)$ given in Proposition~\ref{prop_riem_orth_decomp}, we the following isomorphisms:
\begin{eqnarray}
\id_{\Lambda^2}\otimes \ast:\mathcal{S}(U^*) \xrightarrow{\quad\cong\quad} \Lambda^4(U^*), & \id_{\Lambda^2}\otimes \ast: \Lambda^4(U^*)\xrightarrow{\quad\cong\quad} \mathcal{S}(U^*),\label{firstline}\\
\id_{\Lambda^2}\otimes \ast: \mathcal{H}^S(U^*) \xrightarrow{\quad\cong\quad}\mathcal{W}^A(U^*), &\id_{\Lambda^2}\otimes \ast:\mathcal{W}^A(U^*)\xrightarrow{\quad\cong\quad}\mathcal{H}^S(U^*),\label{secondline}\\
\id_{\Lambda^2}\otimes \ast:\mathcal{W}^S(U^*)\xrightarrow{\quad\cong\quad} \mathcal{W}^S(U^*), & \id_{\Lambda^2}\otimes \ast: \mathcal{H}^A(U^*)\xrightarrow{\quad\cong\quad}\mathcal{H}^A(U^*).\label{thirdline}
\end{eqnarray}
\end{prop}
\pf{
First we note that $\id_{\Lambda^2}\otimes \ast$ is its own inverse since $(\id_{\Lambda^2}\otimes \ast)^2=\id_{\Lambda^2}\otimes\id_{\Lambda^2}$.
Lemma~\ref{lemma_A11}a) states that $\id_{\Lambda^2}\otimes \ast$ maps a basis of $\mathcal{S}(U^*)$ to a basis of $\Lambda^4(U^*)$.
This shows that the two maps in \eqref{firstline} have the given range and are isomorphisms.\medskip

\noindent
It is a classical fact \cite[Chap.\ 13B]{Besse} that, if $\dim(U)=4$, the map $\id_{\Lambda^2}\otimes \ast$ preserves the space $\mathcal{W}^S(U^*)$ of algebraic Weyl tensors and is an isomorphism.\medskip

\noindent
Next, we show that $\id_{\Lambda^2}\otimes \ast$ maps $\mathcal{H}^S(U^*)$ into $\mathcal{W}^A(U^*)$.
Let $h\owedge g\in \mathcal{H}^S(U^*)$, i.e.\ $h\in S^2_0(U^*)$. 
We will verify that $\id_{\Lambda^2}\otimes \ast(h\owedge g)$ is perpendicular to any component of the decomposition from Proposition~\ref{prop_riem_orth_decomp} other than $\mathcal{W}^A(U^*)$.
For $Q\in \mathcal{W}^S(U^*)$ we have $\id_{\Lambda^2}\otimes \ast(Q)\in \mathcal{W}^S(U^*)$ and, therefore,
\[
\langle Q, \id_{\Lambda^2}\otimes \ast(h\owedge g)\rangle =\langle \id_{\Lambda^2}\otimes \ast(Q),h\owedge g \rangle=0
\]
as $\mathcal{W}^S(U^*)$ and $\mathcal{H}^S(U^*)$ are orthogonal.
For $\omega^g\in \Lambda^4(U^*)$ one obtains
\[
\langle \omega^g, \id_{\Lambda^2}\otimes \ast(h\owedge g)\rangle =\langle \id_{\Lambda^2}\otimes \ast(\omega^g),h\owedge g \rangle
=\langle \tfrac14\,g\owedge g,h\owedge g \rangle=0
\]
as $\mathcal{S}(U^*)$ and $\mathcal{H}^S(U^*)$ are perpendicular.
For $k\owedge g\in \mathcal{H}^S(U^*)\oplus \mathcal{H}^A(U^*)$ Lemma~\ref{lemma_A11}b) gives
\[
\langle k\owedge g, \id_{\Lambda^2}\otimes \ast(h\owedge g)\rangle = 0.
\]
Hence $\id_{\Lambda^2}\otimes \ast(\mathcal{H}^S(U^*) )\subset \mathcal{W}^A(U^*)$.
Since $ \dim \mathcal{H}^S(U^*)=\dim\mathcal{W}^A(U^*)=9$ the first map in \eqref{secondline} is an isomorphism and the second one as well.
\medskip

\noindent
Finally, since the decomposition from Proposition~\ref{prop_riem_orth_decomp} is orthogonal and $\id_{\Lambda^2}\otimes \ast$ is an isometry,
the second map in \eqref{thirdline} is also an isomorphism.
\hfill\Box}\medskip

\begin{cor}\label{cor_trace_representation}
For $Q\in \mathcal{H}^A(U^*)\oplus\mathcal{W}^A(U^*)$  we consider the $(2,0)$-tensor
\[
q=-\tfrac12 c\left( \id_{\Lambda^2}\otimes \ast(Q)\right).
\]
Then one has $\tr_g(q)=0$ and $\id_{\Lambda^2}\otimes \ast(Q)=q\owedge g$.
Furthermore we find that  $q\in S^2_0(U^*)$, if $Q\in \mathcal{W}^A(U^*)$, and that $q\in \Lambda^2(U^*)$, if $Q\in\mathcal{H}^A(U^*)$.
\end{cor}
\pf{Proposition~\ref{prop_decompo_Hodge} provides the existence and uniqueness of a tracefree $q$ with $\left( \id_{\Lambda^2}\otimes \ast(Q)\right)=q\owedge g$.
With Lemma~\ref{lemma_kulkarni_trace_and_norm} a) one identifies this $q$ as in the corollary.
\hfill\Box}\medskip

\subsection{The Norm of Curvature Tensors}\label{appendix_A_2}

We consider an $n$-dimensional Riemannian manifold equipped with an orthogonal connection $\nabla$ without Cartan type torsion, thus
there is a vector field $V$ and a  $3$-form $T$ such that \eqref{orthogonal_connection} takes the form
\begin{equation}\label{connection_appendix_Cartan_zero}
\nabla_XY=\nabla^g_XY +\langle X,Y \rangle V- \langle V,Y \rangle X + T(X,Y,\cdot)^\sharp
\end{equation}
for any vector fields $X,Y$.\medskip

\noindent
For a fixed point $p\in M$ we will apply the considerations of section~\ref{appendix_A_1} to the tangent space $T_pM$.
For convenience we introduce the symmetric $(2,0)$-tensors $g^T$, $g^V$, $g^{\nabla V}$ on $T_pM$, for $X,Y\in T_pM$ they are given by
\begin{align*}
g^T(X,Y)&=  \langle X\lrcorner T, Y \lrcorner T\rangle, \\
g^V(X,Y)&=  \langle V,X \rangle \langle V,Y \rangle, \\
g^{\nabla V}(X,Y)&= \langle \nabla^g_X V,Y \rangle +\langle\nabla^g_Y V,X \rangle.
\end{align*}
\begin{lemma}\label{lemma_special_traces}
The traces of the above $(2,0)$-tensors are
\[
\tr_g(g^T)= \|T\|^2,\;\;  \tr_g(g^V)= |V|^2,\;\; \tr_g(g^{\nabla V})=2\divergenz^g(V).
\]
\end{lemma}
\pf{Direct calculation. \hfill \Box}\medskip

\noindent
We note that the anti-symmetric $(2,0)$-tensors $d(V^\flat)$ on $T_pM$ can be written as
\[
d(V^\flat)(X,Y)= \langle \nabla^g_X V,Y \rangle -\langle\nabla^g_Y V,X \rangle.
\]

\noindent
Furthermore, we define the $(4,0)$-tensors $G^T$, $K^{\nabla V}$ and $K^{VT}$ on $T_pM$ by
\begin{align*}
G^T(X,Y,Z,W)&=\langle T(Y,Z,\cdot)^\sharp, T(X,W,\cdot)^\sharp \rangle -\langle T(X,Z,\cdot)^\sharp, T(Y,W,\cdot)^\sharp\rangle, \\
K^{\nabla T}(X,Y,Z,W)&= \left(\nabla^g_XT \right)(Y,Z,W)-\left(\nabla^g_YT \right)(X,Z,W)
-\left(\nabla^g_ZT \right)(X,Y,W)+\left(\nabla^g_WT \right)(X,Y,Z),\\ 
K^{VT}(X,Y,Z,W)&= \langle V,X \rangle \cdot T(Y,Z,W))  -\langle V,Y \rangle\cdot  T(X,Z,W)-\langle V,Z \rangle \cdot T(X,Y,W) +\langle V, W \rangle\cdot  T(X,Y,Z).
\end{align*}
\begin{rem}\label{rem_G_T}
For the Ricci contraction we get $c(G^T)=g^T$. 
Furthermore we see that $G^T\in S^2(\Lambda^2 T_pM^*)$, and therefore $(\id -b)(G^T)\in\mathcal{R}(T^*_pM)$
\end{rem}
\begin{rem}\label{rem_ricci-contraction_and-4-form}
We note that $V^\flat\wedge T$ is a $4$-form, and hence it is in the image of the Bianchi map and its Ricci contraction vanishes.
\end{rem}

\noindent
For $\nabla$, given as in \eqref{connection_appendix_Cartan_zero}, 
the Riemann curvature tensor $\riem$ is defined as in \eqref{def_riem}.
We note that $\riem$ is anti-symmetric in the first and the last pair of its entries.
Yet it is not an algebraic curvature tensor in the sense of \eqref{def_algebraic_curvature_tensor} since in general it does not coincide with its {\it symmetrisation}
\[
\riem^S(X,Y,Z,W)= \tfrac12\left( \riem(X,Y,Z,W)+ \riem(Z,W,X,Y)\right).
\]
The {\it anti-symmetric part} of the Riemann curvature tensor is accordingly defined as
\[
\riem^A(X,Y,Z,W)= \tfrac12\left( \riem(X,Y,Z,W)- \riem(Z,W,X,Y)\right),
\]
which yields the decomposition $\riem=\riem^S +\riem^A$ which is orthogonal with respect to the scalar product of $(4,0)$-tensors.
\begin{lemma}
The components of the Riemann curvature tensor of $\nabla$ are
\begin{align}
\riem^S &= \riem^g -\tfrac12\cdot g^{\nabla V}\owedge g +\tfrac12 |V|^2\cdot g\owedge g - g^V\owedge g
+\tfrac12\cdot dT -V^\flat\wedge T - G^T,\label{eq_symm_riemann} \\
\riem^A &= \tfrac12\cdot K^{\nabla T}-\tfrac12\cdot d(V^\flat) \owedge g +  \left(V\lrcorner T \right) \owedge g+K^{VT}.
\label{eq_antisymm_riemann}
\end{align}
\end{lemma}
\pf{This is verified by a lengthy but still elementary calculation. \hfill \Box}

\begin{cor}\label{cor_A19}
As $\riem^S \in S^2(\Lambda^2T_pM^*)$, it has a component $\riem^S_{\ker} = \left( \id-b\right) (\riem^S) $  in 
the kernel of the Bianchi map  and  a component $\riem^S_{\im} = b(\riem^S)$ in the image:
\begin{align*}
\riem^S_{\ker} &= \riem^g -\tfrac12\cdot g^{\nabla V}\owedge g +\tfrac12 |V|^2\cdot g\owedge g - g^V\owedge g -\left( \id-b\right)(G^T) \in\mathcal{R}(T_pM), \\
\riem^S_{\im} &=\tfrac12\cdot dT -V^\flat\wedge T - b(G^T)\in \Lambda^4 T_pM^*.
\end{align*}
\end{cor}
\pf{We use $\Lambda^4 T_pM^*= S^2(\Lambda^2T_pM^*)\cap \im(b)$ and
Remarks~\ref{rem_automatisch_curvature}, \ref{rem_G_T} and \ref{rem_ricci-contraction_and-4-form}.\hfill \Box}\medskip

\noindent
We can rewrite the definition of the Ricci curvature \eqref{def_ric} as $\ric = c(\riem)$.
Its symmetric and anti-symmetric component are defined as $\ric^S(X,Y)=\tfrac12 \left(\ric(X,Y)+ \ric(Y,X)\right)$ and $\ric^A(X,Y)=\tfrac12 \left(\ric(X,Y)- \ric(Y,X)\right)$.
\begin{rem}
One notices that $\ric^S=c(\riem^S)$ and $\ric^A=c(\riem^A)$.
\end{rem}
\begin{lemma}\label{lemma_a14}
We have
\begin{align*}
\ric^S&= \ric^g+\left(\divergenz^g(V)-(n-2)|V|^2 \right) \, g+(n-2)\,g^V+\tfrac{n-2}{2}\,g^{\nabla V}-g^T,\\
\ric^A&= -\delta T+\tfrac{n-2}{2}\, d(V^\flat)+(4-n)\,(V\lrcorner T),
\end{align*}
where $\delta T$ is the codifferential of $T$, which is given as $\delta T=-\sum_{i=1}^n E_i\lrcorner\, (\nabla^g_{E_i} T)$ for any orthonormal basis $E_1,\ldots,E_n$ (or equivalently $\delta T =-\ast d\ast T$, if $M$ carries an orientation).
\end{lemma}
\pf{The Ricci contractions of \eqref{eq_symm_riemann} and \eqref{eq_antisymm_riemann} are evaluated using Lemma~\ref{lemma_special_traces}.
\hfill\Box}\medskip

\noindent
Applying Lemma~\ref{lemma_special_traces} again, we evaluate the scalar curvature as
\begin{equation}\label{eq_scalar_curvature}
R=  \tr_g(\ric^S)= R^g+2(n-1)\divergenz^g(V) -(n-1)(n-2)|V|^2-\|T\|^2,
\end{equation}
which agrees with the formula given in \cite[Lemma 2.5]{PfaeffleStephan11a}.\medskip

\noindent
For the remainder of this section we assume that $M$ is $4$-dimensional.
\medskip

\noindent
Since $\riem^S_{\ker}$ is an algebraic curvature tensor it possesses a Ricci decomposition as \eqref{eq_Ricci_decomposition}.
Its algebraic Weyl tensor can be identified with the Weyl curvature of the Levi-Civita connection:
\begin{lemma}\label{Weyl_bleibt_Weyl.}
Let $\dim(M)=4$ and let $C^g=\W(\riem^g)$ denote the Weyl curvature of the Levi-Civita connection.
Then we have
\[
C^g =\W(\riem^S_{\ker}).
\]
\end{lemma}
\pf{ In this situation we have
\begin{align*}
\W(\riem^S_{\ker}) &= -\tfrac{1}{12}\, R\,g\owedge g +\tfrac12\,\ric^S\owedge g + \riem^S_{\ker}\\
&= C^g +\left( \tfrac{1}{12}\,\|T\|^2\, g\owedge g-\tfrac12\,g^T\owedge g- ( \id-b)(G^T) \right).
\end{align*}
Setting $C^T=\tfrac{1}{12}\,\|T\|^2\, g\owedge g-\tfrac12\,g^T\owedge g- ( \id-b)(G^T)$ we need to show that $C^T=0$.
We note that $C^T$ is an algebraic Weyl tensor, $C^T=\W( - ( \id-b)(G^T))$. 
(This can easily be verified for the situation $\riem^g\equiv 0$ and $V\equiv 0$ where $\riem^S_{\ker}=- ( \id-b)(G^T)$.)
As the Ricci decomposition is orthogonal we have
\begin{align*}
\|C^T\|^2&= \langle C^T, - ( \id-b)(G^T) \rangle \\
&=  - \langle ( \id-b)C^T,G^T \rangle\\
&= - \langle C^T,G^T\rangle,
\end{align*}
the last equality holds as $C^T\in\ker(b)$.
Finally, we note that \cite[Lemma~A.1]{PfaeffleStephan11a} still holds true if one replaces the Weyl curvature of the Levi-Civita connection by any algebraic Weyl tensor such as $C^T$.
It states that $\langle C^T,G^T\rangle=0 $ if $\dim(M)=4$.
\hfill\Box}\medskip

\begin{cor}\label{cor_a16}
If $\dim(M)=4$  the $(4,0)$-tensor norm of $\riem^S_{\ker}$ is given by
\[
\|\riem^S_{\ker}\|^2 = \|C^g\|^2 +2\,\| \ric^S\|^2 -\tfrac13 R^2.
\]
\end{cor}
\pf{Since $h(\riem^S_{\ker})$ is tracefree Lemma~\ref{lemma_kulkarni_trace_and_norm}b) shows
\begin{align*}
\langle h(\riem^S_{\ker})\owedge g,h(\riem^S_{\ker})\owedge g \rangle &= 8 \langle h(\riem^S_{\ker}),h(\riem^S_{\ker})\rangle \\
&= 8 \langle \ric^S-\tfrac14\,R\, g, \ric^S-\tfrac14\,R\, g\rangle\\
&= 8\,\|\ric^S\|^2- 4\,R\,\langle \ric^S,g\rangle +\tfrac12 \,R^2\, \langle g,g\rangle\\
&= 8\,\|\ric^S\|^2 - 2\,R^2.
\end{align*}
We use the orthogonality of the Ricci decomposition and Lemma~\ref{Weyl_bleibt_Weyl.} to finish the proof. \hfill\Box}\medskip

\begin{lemma}\label{lemma_a17}
If $\dim(M)=4$ the $(4,0)$-tensor norm of $\riem^S_{\im}$ is given by
\[
\|\riem^S_{\im}\|^2 = \tfrac14\,\| dT\|^2 +4\,|V|^2\,\| T\|^2 -12\,\|V\lrcorner T\|^2- 4\,\langle  V\lrcorner dT, T\rangle.
\]
\end{lemma}
\pf{ For any $4$-form $\omega$ the argument which leads to (62) in \cite{PfaeffleStephan11a} shows that
\begin{equation}\label{eq_omega_GT_null}
0=\langle \omega^g,G^T \rangle =\langle b(\omega^g),G^T \rangle =\langle \omega^g,b(G^T) \rangle.
\end{equation}
Hence $\langle dT,b(G^T) \rangle = \langle V^\flat\wedge T,b(G^T) \rangle= \langle b(G^T),b(G^T)\rangle =0$.
From this we get
\[
\|\riem^S_{\im}\|^2 = \tfrac14 \,\|dT\|^2 +\langle V^\flat\wedge T,V^\flat\wedge T\rangle -\langle dT, V^\flat\wedge T \rangle.
\]
Using the definition of the scalar product of $(4,0)$-tensors we find
\begin{align*}
\langle V^\flat\wedge T, V^\flat\wedge T \rangle &=4\,|V|^2\,\| T\|^2 -12\,\|V\lrcorner T\|^2,\\
\langle dT, V^\flat\wedge T \rangle&=  4\,\langle  V\lrcorner dT, T\rangle,
\end{align*}
from which the Lemma follows.
\hfill\Box}

\section{Calculations with Topological Classes}\label{appendix_B}
\subsection{Euler Characteristics}
We consider a $4$-dimensional  compact Riemannian manifold $M$ without boundary.
Let $\nabla$ be an orthogonal connection which takes the form of \eqref{connection_appendix_Cartan_zero}.
\begin{lemma}\label{lemma_b1}
The integral over the square of the norm of the curvature tensor of $\nabla$ is given by
\begin{align*}
\int_M \|\riem\|^2 \,dx& = \int_M \Big(\tfrac13 \,R^2- \tfrac23\,(R^g)^2+2\| \ric^g\|^2 +\|C^g\|^2 +4\| \delta T\|^2 +4\| d(V^\flat)\|^2 \\
& \qquad\qquad +\tfrac12\,\| dT\|^2 +8\,|V|^2\,\| T\|^2 -24\,\|V\lrcorner T\|^2- 8\,\langle  V\lrcorner dT, T\rangle\Big) \,dx
\end{align*}
\end{lemma}
\pf{ The classical formula of the Euler characteristics of $M$ (see \cite{Berger}) states
\begin{equation}\label{eq_euler_1}
8\pi^2\chi(M)=\int_M \left((R^g)^2-4\|\ric^g\|^2+\|\riem^g\|^2 \right) \,dx,
\end{equation}
and one can show (see \cite[Thm.\ 3.4]{PfaeffleStephan11a}) that
\begin{equation}\label{eq_euler_2}
8\pi^2\chi(M)=\int_M \left(R^2-4\|\ric^S\|^2+4\|\ric^A\|^2  +\|\riem^S\|^2- \|\riem^A\|^2 \right) \,dx.
\end{equation}
Hence the difference of the right hand sides of \eqref{eq_euler_2} and \eqref{eq_euler_1} is zero.
By adding this difference and using $\|\riem\|^2=\|\riem^S\|^2+\|\riem^A\|^2 $ we obtain
\begin{equation*}
\int_M \|\riem\|^2 \,dx=\int_M\left( R^2 -(R^g)^2- 4\|\ric^S\|^2+4\|\ric^A\|^2 + 4\|\ric^g\|^2 +2\|\riem^S\|^2 -\|\riem^g\|^2 
\right)\,dx.
\end{equation*}
The Ricci decomposition yields $\| \riem^g\|^2 = \|C^g\|^2 + 2\|\ric^g\|^2-\tfrac13 (R^g)^2$.
Furthermore we know $\|\riem^S\|^2 = \|\riem^S_{\ker}\|^2+ \|\riem^S_{\im}\|^2 $.
Since $\int_M \langle \delta T, dV^\flat \rangle \, dx = \int_M \langle T, d dV^\flat\rangle\, dx =0$ we have
\[
\int_M  \|\ric^A\|^2 \,dx= \int_M\left( \langle\delta T, \delta T\rangle + \langle dV^\flat,dV^\flat \rangle-2 \langle\delta T, dV^\flat\rangle  \right) \,dx = \int_M \left( \|\delta T \|^2 + \|dV^\flat \|^2\right)\,dx
\]
Then we are able to impose the explicit formulas for all occuring terms (Lemma~\ref{lemma_a14}, Corollary~\ref{cor_a16} and  Lemma~\ref{lemma_a17}),
which finishes the proof.
\hfill\Box}

\subsection{First Pontryagin Class}\label{appendix_B2}

We consider a $4$-dimensional compact oriented Riemannian manifold $M$ without boundary.
Let $\nabla$ be an orthogonal connection which takes the form of \eqref{connection_appendix_Cartan_zero}.
Let $E_1,\ldots, E_4$ be a local orthonormal frame of the tangent bundle $TM$.
The curvature $2$-forms w.r.t.\ this frame are defined as
\[
\Riem_{ij}=\sum_{k,\ell} \riem(E_1,E_j,E_k,E_\ell) E_k^\flat\wedge E_\ell^\flat
\]
for $i,j=1,\ldots,4$.
By Chern-Weil theory (see e.g.\ \cite[Chap.~XII.4]{KN69} and \cite[Chap.~13B]{Besse}) the $4$-form $\sum_{i,j} \Riem_{ij}\wedge\Riem_{ij}$ is independent of the choice of the frame and integrates to the {\it first Pontryagin class} of $M$ up to a prefactor:
\begin{align}
p_1(M)&=-\tfrac{1}{8\pi^2}\int_M \sum_{i,j} \Riem_{ij}\wedge\Riem_{ij}\nonumber \\
&= -\tfrac{1}{16\pi^2}\int_M \sum_{i,j} \langle \Riem_{ij},\ast \Riem_{ij}\rangle \,dx\nonumber \\
&= -\tfrac{1}{16\pi^2}\int_M \langle \riem, \id_{\Lambda^2}\otimes \ast(\riem) \rangle \, dx .\label{eq_pontryagin_class}
\end{align}
Now we decompose the curvature tensor of $\nabla$ according Proposition~\ref{prop_riem_orth_decomp}:
\begin{equation}\label{eq_riem_decompo}
\riem= -\tfrac{1}{24}R\, g\owedge g -\tfrac12\,h^S \owedge g +C^g+ b(\riem^S)-\tfrac12\ric^A\owedge g +(\riem^A +\tfrac12 \ric^A\owedge g)
\end{equation}
where we have inserted Lemma~\ref{Weyl_bleibt_Weyl.} and have denoted $h^S=h(\riem^S_{\ker})=\ric^S-\tfrac14\,R\,g$.\medskip

\noindent
We use Proposition~\ref{prop_decompo_Hodge} and the orthogonality of the decomposition in Proposition~\ref{prop_riem_orth_decomp} and obtain
\begin{align}
\langle \riem, \id_{\Lambda^2}\otimes \ast(\riem)\rangle 
=&
\langle C^g, \id_{\Lambda^2}\otimes \ast(C^g)\rangle
+\tfrac14\langle \ric^A\owedge g, \id_{\Lambda^2}\otimes \ast(\ric^A\owedge g)\rangle\nonumber\\
&-\tfrac{R}{12}\langle \riem^S_{\im}, \id_{\Lambda^2}\otimes \ast(g\owedge g)\rangle
-\langle \riem^A +\tfrac12 \ric^A\owedge g, \id_{\Lambda^2}\otimes \ast(h^S\owedge g)\rangle.\label{riem_Hodge_Pontryagin}
\end{align}
To the authors' knowledge Baekler and Hehl were the first to notice that the density of the first Pontryagin class can be put in such a form, see \cite[Section 3]{BH11}.
\begin{rem}\label{rem_B2}
The integral over the second summand of the right hand side of \eqref{riem_Hodge_Pontryagin} is 
\[
\tfrac14\int_M \langle \ric^A\owedge g, \id_{\Lambda^2}\otimes \ast(\ric^A\owedge g)\rangle\,dx = 4\int_M \langle d\ast T, dV^\flat\rangle\, dx.
\]
\end{rem}
\pf{ By Lemma~\ref{lemma_A11}c) we have $\langle \ric^A\owedge g, \id_{\Lambda^2}\otimes \ast(\ric^A\owedge g)\rangle
=8\langle \ric^A, \ast\ric^A\rangle$.
We observe that $\delta \alpha=-\ast d\ast \alpha$ for any $k$-form $\alpha$, which gives that $\int_M \langle \delta T,\ast\delta T\rangle\,dx = 0$ and $\int_M \langle dV^\flat, \ast dV^\flat\rangle\,dx = 0$.
This leads to
\[
\int_M \langle \ric^A, \ast\ric^A\rangle \,dx = \int_M \langle \delta T- dV^\flat , \ast\delta T  -\ast dV^\flat\rangle \,dx =
2 \int_M \langle d\ast T, dV^\flat\rangle\, dx,
\]
from which the claim follows.
\hfill\Box}

\begin{rem}\label{rem_B3}
For the third summand of the right hand side of \eqref{riem_Hodge_Pontryagin} we get
\[
-\tfrac{R}{12}\langle \riem^S_{\im}, \id_{\Lambda^2}\otimes \ast(g\owedge g)\rangle = -\tfrac{R}{6} \langle dT, \omega^g\rangle + \tfrac{4R}{3}\langle
T,\ast V^\flat\rangle. 
\]
\end{rem}
\pf{
From Lemma~\ref{lemma_A11}a) and Corollary~\ref{cor_A19}  we obtain 
\begin{align*}
\langle \riem^S_{\im}, \id_{\Lambda^2}\otimes \ast(g\owedge g)\rangle &= 4\langle \riem^S_{\im}, \omega^g \rangle \\
&= 2\langle dT, \omega^g \rangle  - 4 \langle V^\flat\wedge T, \omega^g  \rangle -4\langle b(G^T),\omega^g \rangle.
\end{align*}
The last term is zero due to \eqref{eq_omega_GT_null}, and from $\langle V^\flat\wedge T, \omega^g  \rangle = 4 \langle
T,\ast V^\flat\rangle$ the claim follows.
\hfill\Box}

\begin{rem}\label{rem_B4}
The integral over the fourth summand of the right hand side of \eqref{riem_Hodge_Pontryagin} is 
\[
\int_M \langle \riem^A +\tfrac12 \ric^A\owedge g, \id_{\Lambda^2}\otimes \ast(h^S\owedge g)\rangle \,dx =
\int_M \left( 4\langle d\ast T, dV^\flat\rangle+ \tfrac{4R}{3}\langle T,\ast V^\flat\rangle\right)\,dx 
-\int_M 4R\,dT
\]
\end{rem}
\pf{
Since the Pontryagin class is a topological invariant we get from \eqref{eq_pontryagin_class} and 
\eqref{riem_Hodge_Pontryagin} that
\[
-16 \pi^2 p_1(M)= \int_M \langle \riem, \id_{\Lambda^2}\otimes \ast(\riem) \rangle \, dx
=  \int_M \langle C^g, \id_{\Lambda^2}\otimes \ast(C^g)\rangle\, dx.
\]
Therefore the last three terms on the right hand side of \eqref{eq_pontryagin_class} integrate to zero.
We note that $\langle\eta, \omega^g\rangle\omega^g=24\,\eta$ for any $4$-form $\eta$,
and with Remarks~\ref{rem_B2} and \ref{rem_B3} we obtain the claim.
\hfill\Box}

\section{Some Traces of the Potential}\label{appendix_C}

In order to evaluate the Seeley-deWitt coefficients for $D^*D$ we need to compute several traces involving the potential that comes from the Lichnerowicz formula \eqref{Lichnerowicz}.
We only consider the case of a $4$-dimensional Riemannian spin manifold $M$.
Comparing \eqref{Lichnerowicz} with the Bochner formula \eqref{Bochner_formula} gives us the potential
\begin{equation}\label{eq_def_potential}
E = \left( -\tfrac14 R^g  +\tfrac34\|T\|^2
\; - \tfrac{3}{2}\divergenz^g(V) +\tfrac92\,|V|^2\right) \id_\Sigma - \tfrac32 dT
 \; -9\,T\cdot V -9\,(V\lrcorner T),
\end{equation}
which acts as an endomorphism of the spinor bundle $\Sigma M$.
We also need the square of the potential:
\begin{align}
E^2 &= \left( -\tfrac14 R^g  +\tfrac34\|T\|^2 \; - \tfrac{3}{2}\divergenz^g(V) +\tfrac92\,|V|^2\right)^2 \id_\Sigma
\nonumber\\
&\quad +2 \left( \tfrac14 R^g  -\tfrac34\|T\|^2 \; + \tfrac{3}{2}\divergenz^g(V) -\tfrac92\,|V|^2\right) \left(\tfrac32 dT +9\,T\cdot V +9\,(V\lrcorner T) \right)
\nonumber\\
&\quad +\tfrac94\, dT\cdot dT+ 81\,\left(T\cdot V + (V\lrcorner T) \right)^2
\nonumber\\
&\quad +\tfrac{27}{2}\left(dT \cdot T\cdot V +T\cdot V \cdot dT\right) + \tfrac{27}{2} \left( dT\cdot (V\lrcorner T)+ (V\lrcorner T) \cdot dT\right).
\label{eq_def_potential_quadrat}
\end{align}
This leads us to the following lemma:

\begin{lemma}\label{lemma_C1}
The traces of $E$ and $E^2$ taken over the spinor bundle are
\begin{align}
\tr_\Sigma(E)&= -R^g  +3\,\|T\|^2
\; - 6\,\divergenz^g(V) +18\,|V|^2\label{eq_trace_E}\\
\tr_\Sigma(E^2)&=4\,\left( -\tfrac14 R^g  +\tfrac34\|T\|^2 \; - \tfrac{3}{2}\divergenz^g(V) +\tfrac92\,|V|^2\right)^2\nonumber\\
&\qquad + \tfrac38\,\|dT\|^2 -162\,\|V\lrcorner T \|^2+ 54\, |V|^2\,\|T\|^2 -18\,\langle V\lrcorner dT, T\rangle
\label{eq_trace_E_squared}
\end{align}
\end{lemma}
\pf{ As $\Sigma M$ has rank $4$ we get $\tr_\Sigma(\id_\Sigma)=4$ and \eqref{eq_trace_E} follows from \eqref{eq_def_potential} immediately.\medskip

\noindent
In order to conclude \eqref{eq_trace_E_squared} from \eqref{eq_def_potential_quadrat} we take into account that in the Clifford algebra only elements of degree zero (i.e.\ elements proportional to the identity) contribute in the trace.
Furthermore we only use cyclicity of the trace and the relations for any tangent vector $V$ and any $k$-forms $\alpha$, $\beta$:
\begin{eqnarray*}
V\cdot \alpha +(-1)^{k+1} \alpha\cdot V&=& -2\,(V\lrcorner\alpha),\\
\tr_\Sigma(\alpha\cdot\beta)&=& \tfrac{4}{k!}\,(-1)^{k(k+1)/{2}}\,\langle \alpha,\beta\rangle,
\end{eqnarray*}
and $V\cdot\alpha = -\,(V\lrcorner\alpha)$ if $\alpha$ is a $4$-form.
\hfill\Box}
\medskip

\noindent
We denote the Riemannian volume form on $M$ by $\omega^g$ and we evaluate $\tr_\Sigma(E\,\omega^g)$ and $\tr_\Sigma(E^2\,\omega^g)$.

\begin{lemma}\label{lemma_C2}
The traces of $E\,\omega^g$ and $E^2\,\omega^g$ taken over the spinor bundle are
\begin{align}
\tr_\Sigma(E\,\omega^g)&= -\tfrac14\, \langle dT,\omega^g\rangle + 6\,\langle T,\ast V^\flat \rangle \label{eq_trace_E_gamma}\\
\tr_\Sigma(E^2\,\omega^g)&= \left( \tfrac14 R^g  -\tfrac34\|T\|^2 \; + \tfrac{3}{2}\divergenz^g(V) -\tfrac92\,|V|^2\right)\,\left(\tfrac12\, \langle dT,\omega^g\rangle - 12\,\langle T,\ast V^\flat \rangle \right) 
\label{eq_trace_E_squared_gamma}
\end{align}
\end{lemma}
\pf{ Apart from the ingredients mentioned in the proof of Lemma~\ref{lemma_C1} we only need that $V\cdot\omega^g=-\ast V^\flat$ for any tangent vector $V$ in order to prove \eqref{eq_trace_E_gamma}.\medskip

\noindent
For the proof of \eqref{eq_trace_E_squared_gamma} we use again that Clifford elements of non-zero degree have trace zero.
Therefore, we have
\begin{align*}
0&=\tr_\Sigma\left(\left( -\tfrac14 R^g  +\tfrac34\|T\|^2 \; - \tfrac{3}{2}\divergenz^g(V) +\tfrac92\,|V|^2\right)^2 \id_\Sigma\,\omega^g\right),\\
0&=\tr_\Sigma\left(\tfrac94\, dT\cdot dT
 \,\omega^g\right),\\
0&=\tr_\Sigma\left(\tfrac{27}{2}\left(dT \cdot T\cdot V +T\cdot V \cdot dT\right)\omega^g + \tfrac{27}{2} \left( dT\cdot (V\lrcorner T)+ (V\lrcorner T) \cdot dT\right) \omega^g\right).
\end{align*}
Since $T\cdot V+(V\lrcorner T)=\tfrac12(T\cdot V -V\cdot T)$ as endomorphisms of spinors we get
\[
\tr_\Sigma
\left(\left(T\cdot V+(V\lrcorner T )\right)^2 \omega^g
\right)
= \tfrac14 \tr_\Sigma\left(\left(T\cdot V -V\cdot T\right)^2 \omega^g\right)
=0
\]
when we take into account that $V\,\omega^g= -\omega^g\, V$ and that the trace is cyclic.
Therefore the only term in $E^2\,\omega^g$ that contributes to the trace is
\[
2 \left( \tfrac14 R^g  -\tfrac34\|T\|^2 \; + \tfrac{3}{2}\divergenz^g(V) -\tfrac92\,|V|^2\right) \left(\tfrac32 dT +9\,T\cdot V +9\,(V\lrcorner T) \right)\,\omega^g.
\]
Its trace is evaluated in the same way as it is done in the proof of \eqref{eq_trace_E_gamma}.
\hfill \Box}\medskip

\noindent
The scalar curvature of the modified connection $\widetilde{\nabla}$ given in \eqref{def_modified_connection} is
\[
\widetilde{R}=R^g+18\divergenz^g(V)-54|V|^2-9\|T \|^2.
\]
We recall that the Holst term $C_H$ computed from the connection $\nabla$ (defined originally in \cite{Holst} or as in (7) in \cite{PfaeffleStephan11b}) is given by
\[
C_H=6\, dT-2\langle T,\ast V^\flat \rangle\,\omega^g -\tfrac12\left(\|S_+\|^2 -\|S_-\|^2\right)\omega^g,
\]
compare Prop.~2.3 in \cite{PfaeffleStephan11b} and its Erratum.
For the modified connection $\widetilde{\nabla}$ the Holst term takes the form
\[
\widetilde{C}_H=18\,dT-18\,\langle T,\ast V^\flat \rangle\,\omega^g.
\]
Using these notations we can rewrite Lemma~\ref{lemma_C1} and Lemma~\ref{lemma_C2}:
\begin{prop}
For the traces taken over the spinor bundle one obtains
\begin{align}
\tr_\Sigma(E)&= -\tfrac23 R^g-\tfrac13\widetilde{R},\label{eq_E_trace}\\
\tr_\Sigma(E^2)&= \tfrac{1}{36}\left(2R^g+\widetilde{R} \right)^2 
+ \tfrac38\,\|dT\|^2 -162\,\|V\lrcorner T \|^2+ 54\, |V|^2\,\|T\|^2 -18\,\langle V\lrcorner dT, T\rangle,
\label{eq_E_squared_trace}\\
\tr_\Sigma(E\omega^g)&= -\tfrac{1}{72}\,\langle \widetilde{C}_H,\omega^g\rangle,\label{eq_E_omega_trace}\\
\tr_\Sigma(E^2\omega^g)&= \tfrac{1}{432}\,\left(2R^g+\widetilde{R} \right)\,\langle \widetilde{C}_H,\omega^g\rangle.
\label{eq_E_squared_omega_trace}
\end{align}

\end{prop}

%
%

\end{document}